\documentclass[useAMS,usenatbib,usegraphicx]{mn2e}
\usepackage{amsmath,mathtools,amsfonts,amssymb,float}
\usepackage[usenames,dvipsnames]{xcolor}
\usepackage{ulem}

\def\Msun{\hbox{$\rm\, M_{\odot}$}}

\newcommand{\twoc}{_{\textrm{200c}}}

\title [Mass and environmental quenching, conformity and clustering] {Galaxy formation in
  the {\it Planck} cosmology - IV. Mass and environmental quenching,
  conformity and clustering.}

\author[Henriques et al.]  {Bruno
  M. B. Henriques$^{1,2}$\thanks{E-mail:bhenriques@mpa-garching.mpg.de},
  Simon D. M. White$^{2}$, Peter A. Thomas$^{3}$, \newauthor
  Raul E. Angulo$^{4}$,  Qi Guo$^{5}$, Gerard Lemson$^{2,6}$, Wenting Wang$^{7}$\vspace{0.4cm}\\
  {}$^{1}$Institute for Astronomy, ETH Zurich, CH-8093 Zurich, Switzerland\\
  {}$^{2}$Max-Planck-Institut f\"ur Astrophysik, Karl-Schwarzschild-Str. 1, 85741 Garching b. M\"unchen, Germany\\
  {}$^{3}$Astronomy Centre, University of Sussex, Falmer, Brighton BN1
  9QH, United Kingdom\\
 {}$^{4}$Centro de Estudios de F\'isica del Cosmos de Arag\'on, Plaza
  San Juan 1, Planta-2, 44001, Teruel, Spain\\ 
  {}$^{5}$Partner Group of the Max-Planck-Institut f\"ur Astrophysik, National Astronomical Observatories, Chinese Academy of Sciences, \\
  ~Beijing, 100012, China\\
  {}$^{6}$Department of Physics and Astronomy, The Johns Hopkins University, Baltimore, MD 21218, USA\\
{}$^{7}$Department of Physics, Institute for Computational Cosmology, University of Durham, South Road, Durham DH1 3LE\\
}

\begin{document}

\date{Submitted to MNRAS}

\pagerange{\pageref{firstpage}--\pageref{lastpage}} \pubyear{2012}

\maketitle

\label{firstpage}

\begin{abstract}
  We study the quenching of star formation as a function of redshift,
  environment and stellar mass in the galaxy formation simulations of
  \citet{Henriques2015}, which implement an updated version of the
  Munich semi-analytic model ({\small L-GALAXIES}) on the two Millennium Simulations after
  scaling to a Planck cosmology.  In this model massive galaxies are
  quenched by AGN feedback depending on both black hole and hot gas
  mass, and hence indirectly on stellar mass. In addition, satellite
  galaxies of any mass can be quenched by ram-pressure or tidal
  stripping of gas and through the suppression of gaseous infall. This
  combination of processes produces quenching efficiencies which
  depend on stellar mass, host halo mass, environment density,
  distance to group centre and group central galaxy properties in ways
  which agree qualitatively with observation.  Some discrepancies
  remain in dense regions and close to group centres, where quenching
  still seems too efficient. In addition, although the mean stellar age of massive galaxies 
  agrees with observation, the assumed AGN feedback model allows too much ongoing 
  star formation at late times. The fact that both AGN feedback and environmental effects
  are stronger in higher density environments leads to a correlation
  between the quenching of central and satellite galaxies which
  roughly reproduces observed conformity trends inside haloes.
 \end{abstract}


\section{Introduction}
\label{sec:intro}

Ever since galaxy morphologies were first systematically classified it
has been clear that the present population can be be split into active
systems where star formation occurs at a significant rate and passive
systems where it is essentially absent \citep[e.g.][]{Hubble1936}. The
first representative surveys showed that the two classes contribute
comparably to the total cosmic mass density in stars, but tend to
inhabit different environments, with passive galaxies being found
preferentially in dense regions \citep{Davis1976, Dressler1980}.  The
very large and well characterised samples provided by the Sloan
Digital Sky Survey made clear that there is a characteristic stellar
mass $\sim 10^{10.5}M_\odot$ above which galaxies are predominantly
passive and below which they are predominantly star-forming
\citep{Kauffmann2003a}. This mass also seems to separate galaxies in
which central supermassive black holes and the associated nuclear
activity play a major role from galaxies in which they are not
normally present \citep{Kauffmann2003b}.

The passive nature and old stellar populations of massive galaxies led
most early modellers to assume that all but the smallest galaxies
assembled at high redshift \citep[e.g.][]{Peebles1988}.  The discovery
that the cosmic star formation rate density rises rapidly with
redshift to a peak at $z\sim2$ \citep{Lilly1996,Madau1996} thus came
as a surprise, supporting the earlier and initially unpopular
suggestion that galaxy formation should be viewed as a process rather
than an event, with cosmic star formation rates peaking over a broad
and relatively late redshift range \citep{White1989}.  Large
observational surveys have now firmly established this behaviour,
showing the median formation redshift of present-day stars to be
$z\sim 1.3$ \citep[e.g.][]{Hopkins2006, Wilkins2008, Karim2011,
  Forster_Schreiber2014}. At each epoch the typical star-formation rate of
active galaxies is approximately proportional to stellar mass, but the
ratio of the two, the ``specific star formation rate'' (SSFR)
increases strongly with redshift \citep{Elbaz2007, Daddi2007,
  Noeske2007, Karim2011}. Such surveys also allow the observed
population to be split into active and passive systems at each
redshift, showing the fraction of passive systems to be much lower at
early times than it is today \citep{Bundy2005, Faber2007, Ilbert2010,
  Pozzetti2010, Muzzin2013, Ilbert2013}.

In hierarchical, dark matter dominated models of structure formation,
the accretion of baryons onto haloes is expected to follow that of the
dark matter, and in this case strong feedback appears necessary to
explain why only a small fraction of all cosmic baryons have been
converted into stars.  This issue was identified long ago using
semi-analytic models \citep{White1978, White1991} and is closely
related to the issue of why the most massive galaxies are currently
almost all passive \citep{Benson2003}. Until recently, feedback
effects in numerical simulations of galaxy formation were generally
too weak to reduce star formation efficiencies to the observationally
required levels \citep[e.g.][]{Guo2010}. Although recent simulations
do approximately reproduce star formation rates in active galaxies,
they still struggle to produce sufficiently passive galaxies at high
mass \citep{Vogelsberger2014, Schaye2015}.  Thus, while the regulation
of star formation by feedback now appears to be adequately
represented, its ``quenching'' at high mass is not.

Two distinct types of process have been proposed to quench star
formation. One, purely internal to the galaxy and active primarily at
high mass, is usually identified as feedback from an active galactic
nucleus (AGN) although the evidence for this is largely circumstantial
\citep{Benson2003, Croton2006, Bower2006, Menci2006, Somerville2008,
  Schaye2010, Vogelsberger2014, Schaye2015}. The other reflects
interactions with the larger-scale environment, typically satellite
interactions with the tidal field and intracluster medium of the host
halo, as well as collisions with other satellites \citep{Gunn1976,
Larson1980, Moore1996}. The implementation of such processes in
semi-analytic galaxy formation models has generally combined with the
assumed truncation of gas accretion to produce excessive quenching of
low mass satellites \citep{Weinmann2006a, Henriques2008, Guo2011,
  Weinmann2012}. Better agreement with observation is found in recent
hydrodynamic simulations \citep{Bahe2015, Sales2015}. In the context
of the semi-analytic models, AGN feedback acts earlier and more
strongly on higher mass galaxies, with the result that their stellar
populations are older at late times. Low mass galaxies are only
quenched if they become satellites and, since the satellite fraction
never exceeds $\sim 50\%$, many such galaxies remain star-forming
until the present day.

A more detailed separation of the two kinds of effect has become
possible as observed samples have become large enough to estimate the
efficiency of quenching as a function of galaxy properties such as
halo, stellar and bulge mass, AGN activity and various measures of
environment \citep{Kauffmann2004, Baldry2006, Wang2008, Peng2010,
  Wetzel2012, Bluck2014, Knobel2015, Woo2015}. The observational trends
are often summarised in terms of increased probabilities of being
quenched at higher stellar mass and at higher environmental density,
with the two effects acting approximately independently. In a recent
paper, \citet{Terrazas2016b} showed that current observational data 
indicate that galaxies of given stellar mass host systematically more massive 
black holes if they are quenched than if they are actively forming stars. 
Of the simulations they analysed, including the \citealt{Henriques2015}
model used in this work, only those having AGN-feedback-dependent
quenching reproduced this observational trend, suggesting the possible importance of this process. 
It should however be noted that, if the normalisation of the black hole
mass-stellar mass relation decreases towards late times, this might simply 
reflect the early quenching of massive galaxies \citep{Caplar2015}.

In \citet{Henriques2015}, hereafter Paper I, we updated the {\small L-GALAXIES}
model of galaxy formation to improve its representation
of the observed evolution of galaxy abundance and quenched fraction as
a function of stellar mass over the redshift range $0<z\leq 3$. We
also scaled the two Millennium Simulations on which the model is
implemented to the cosmological parameters of \citet{Planck2014}.  In
the present paper we test how well this model accounts for quenching
as a function of environment. In particular, we study the
quenching-stellar mass relation separately for central and satellite
galaxies, as well as its dependence on host halo mass and halocentric
distance. We also study galaxy autocorrelations as a function of
stellar mass both for star-forming and for passive galaxies.  In
all cases we take care that the definitions of galaxy samples and of
environmental measures are as close as possible to those used in the
observational studies with which we compare.  Finally, we test how
well our model reproduces observed conformity effects, the tendency
for the SSFR of satellites to correlate with that of the central
galaxy they orbit.

Our paper is organised as follows. Section~\ref{sec:munich_model}
summarises the most relevant of the modifications introduced in the
galaxy formation model by \citet{Henriques2015} showing how they
affect the abundances of satellite and central galaxies as a function
of star-formation activity and redshift. Section~\ref{sec:agn}
focusses on how well the AGN feedback model reproduces the observed
quenching of high mass galaxies as a function of stellar and halo
mass. Section~\ref{sec:env} compares recent observational data with
predictions for quenched fractions as a function of environment as
traced by host halo mass, by distance to the fifth nearest neighbour
and by distance to the halo centre, while section~\ref{sec:clustering}
studies autocorrelation functions for passive and star-forming
galaxies as a function of their stellar mass.
Section~\ref{sec:conformity} looks at how the combined effects of AGN
feedback and environment give rise to conformity between central and
satellite galaxies. Finally, we draw our conclusions in
section~\ref{sec:conclusions}.


\section{The Munich Model of Galaxy Formation - {\small L-GALAXIES}}
\label{sec:munich_model}

The {\small L-GALAXIES} model of galaxy formation includes prescriptions for a wide
variety of physical processes in an attempt simulate all the phenomena
that shape the galaxy population.  A full description of the current
prescriptions, and in particular of changes made for this series of
papers with respect to earlier work, was presented in the
supplementary material to Paper I.  These changes primarily affect low
mass galaxies, $8.0 \leq \log_{10} (M_*[\Msun]) \leq 9.5$, enhancing their
star formation at later times by: (i) providing additional fuel
through delayed reincorporation of gas previously ejected by SN (this
mainly affects central galaxies); (ii) defining a threshold halo mass
for ram-pressure stripping to be effective; (iii) lowering the
threshold for star formation (these two changes mainly affect
satellite galaxies). The first modification, related to the reincorporation
of gas ejected from SN feedback, is motivated by the need to decouple 
the accretion of baryons into galaxies from that onto haloes and ensure
that enough gas is available to cool in low mass galaxies at late times. 
A similar result was found in the hydro simulations of \citet{Oppenheimer2008} and \citet{Oppenheimer2010}.
The ram-pressure stripping threshold was introduced to avoid that the 
excessive hot gas content of intermediate mass groups in
our model produces an effect on satellite galaxies that incorrectly 
dominates over any other environmental process. We are currently 
testing if an AGN feedback wind model is capable
of ejecting the gas in order to eliminate the need for this threshold 
(\citealt{Bower2008}; Fournier et al. in prep).

A modification to the AGN feedback model ensures
that intermediate mass galaxies ($\log_{10} (M_*[\Msun]) \sim 10.5$) have
significant star formation below $z=2$, but are predominantly quenched
by $z=0$. Specifically, the heating rate is taken to be 
$\dot{E}\propto M_{\rm{BH}}M_{\rm{hot}}$ rather
than $\dot{E}\propto M_{\rm{BH}}M_{\rm{hot}}H(z)$ (the form used in
  \citealt{Croton2006} and \citealt{Guo2011}). Although the latter is closer
  to what is expected from Bondi accretion, the details of gas infall
onto black holes remain uncertain and must be inferred from their
impact on the properties of massive galaxies.
The large parameter space of the current model has been
fully explored using the MCMC method introduced in
\citet{Henriques2009} and \citet{Henriques2010}. Finally, the model is
built on the Millennium simulations after scaling to the first-year
{\it Planck} cosmology (\citealt{Planck2014}; 
$\sigma_8=0.829$, $H_0=67.3\;\rm{km}\;\rm{s}^{-1}\rm{Mpc}^{-1}$,
$\Omega_{\Lambda}=0.685$, $\Omega_{\rm{m}}=0.315$,
$\Omega_{\rm{b}}=0.0487$, $f_{\rm{b}}=0.155$ and $n=0.96$) and adopts the
\citet{Maraston2005} stellar populations originally introduced in the
Munich model by \citet{Henriques2011, Henriques2012}. The change in
stellar population model has negligible impact at the relatively low
redshifts considered in this paper. Throughout the paper, the 
Millennium-II Simulation is used for $\log_{10} (M_*[\Msun]) \leq 9.5$
 and the Millennium Simulation for higher stellar masses. Above this mass cut, 
 properties of galaxies are nearly identical in the two simulations. This includes
 the properties of their super-massive black holes which are particularly 
 relevant for the quenching phenomena discussed in the present work and for which
 numerical convergence is crucial \citep{Angulo2014}.

\subsection{Quenching Mechanisms}
\label{subsec:quench}

This paper focuses primarily on the processes that shut down star
formation and result in evolution of the actively star-forming and
quenched populations of galaxies. In our model, two distinct processes
can cause a galaxy to move from the active to the passive population:
feedback from a central black hole that grows to large mass while
surrounded by a hot gas atmosphere, or becoming a satellite of a larger
companion. In the following subsections we describe the modifications
introduced in Paper I that are particularly relevant to these
processes.  We refer the reader to the original paper for a full
description of the galaxy formation modelling. It should be noted that
although the fraction of passive galaxies increases steadily with
cosmic time, passive galaxies can (and often do) return at least
temporarily to the star-forming population if they accrete new gas
(for example, through a merger).

\subsubsection{Quenching by radio mode feedback}
\label{subsec:radio_quench}

As explained in Appendix\;A10 of Paper I, two major processes increase
the mass of black holes in our model: the quasar and radio accretion
modes. The quasar mode produces no feedback, but is responsible for
the majority of black hole growth as cold gas is accreted during
mergers. Radio-mode feedback results from hot gas accretion onto the
black hole and is assumed to increase strongly with both black hole
mass and hot gas mass. Massive galaxies typically host large black
holes and are surrounded by massive hot gas halos, so feedback is
strong and eliminates further condensation of hot gas onto the
galaxy. Star formation then exhausts the remaining cold gas and is
quenched.  Although star formation can occur subsequently if merging
satellites bring in new cold gas, these massive galaxies live in
massive haloes and their satellites are typically gas-poor as a result
of tidal and ram-pressure stripping. As a result, most massive
galaxies are quenched and their stellar masses grow primarily through
merging.

The model for radio mode feedback in Paper I differs from that in
\citet{Guo2011} and earlier papers by having a factor of $H(z)$ removed. This
enhances feedback at later times and ensures that $M_{\star}$ galaxies
are predominantly quenched by $z=0$ despite forming a significant
fraction of their stars at $z<1$. The heating from feedback is taken
to be proportional to the radio-mode accretion rate, which is
assumed to be

\begin{equation} \label{eq:radio1}
  \dot{M}_{\rm{BH}}=k_{\rm{AGN}}
  \left(\frac{M_{\rm{hot}}}{10^{11}\Msun}\right)\left(\frac{M_{\rm{BH}}}{10^8\Msun}\right),
\end{equation}
while the black hole mass itself is determined by quasar mode accretion in
merger events according to:
\begin{equation} \label{eq:quasar}
\Delta M_{\rm BH,Q}=\frac{f_{\rm BH}(M_{\rm sat}/M_{\rm
    cen})\,M_{\rm cold}}{1+(V_{\rm{BH}}/V_{\twoc})^2}.
\end{equation}
In these formulae $k_{\rm{AGN}}$, $f_{\rm BH}$ and $V_{\rm{BH}}$ are
treated as free parameters.  As a result of these assumptions,
galaxies are more likely to be passive for larger values of black hole
mass ($M_{\rm{BH}}$), hot gas mass ($M_{\rm{hot}}$), halo virial
velocity ($V_{\twoc}$) and in environments where the typical mass
ratio between satellites and centrals ($M_{\rm sat}/M_{\rm cen}$) is
larger. The correlations between these and other galaxy properties,
for example, bulge, stellar and halo mass, naturally induce further
correlations between galaxy structure and quenching. For a complete
description of such relations for Milky Way mass systems see
\citet{Terrazas2016a}.

\subsubsection{Environmental quenching}
\label{subsec:env_quench}

Low-mass galaxies are barely affected by their relatively small
central black holes. On the other hand, both SN feedback and
environment play a crucial role in shaping their properties. The
modification of gas reincorporation introduced in
\citet{Henriques2013} results in material ejected by SN feedback
returning later than in previous models. As a result, star formation is
significantly delayed in isolated low-mass galaxies ($8.0 \leq 
\log_{10} (M_*[\Msun]) \leq 9.5$) peaking at $z < 2$ and continuing at a
substantial rate down to $z=0$.  If such galaxies fall into a group or
cluster, their subsequent evolution changes dramatically. They cease
accreting from the intergalactic medium, the hot gas they do have is
progressively removed by tidal and ram-pressure forces, and the gas
ejected by supernova is permanently lost to the intracluster medium.

This combination of environmental processes has a strong impact on
satellite galaxies and leads to the complete shut-down of star
formation in a significant fraction of them. The overall impact was
too large in previous versions of our model
\citep{Weinmann2006b, Henriques2008, Guo2011}. Hot gas reservoirs were
removed rapidly from satellites and star formation then consumed the
remaining gas until the assumed surface density threshold was
reached. This quenched star formation in the majority of satellite
galaxies, even when they retained significant amounts of cold gas.  In
the new model, ram-pressure is assumed to strip hot gas only in haloes
above a minimum mass $\sim 10^{14}M_\odot$, and the gas surface
density threshold for star formation is roughly halved from its
earlier value. Combined, these modifications ensure that more hot gas
remains in satellites and that, once cooled, more of it is converted
into stars -- delayed reincorporation of ejected gas provides
enough fuel to low-mass central galaxies to drive star formation at
late times, and our reductions in the strength of ram-pressure
stripping and in the star formation threshold ensure that this is also
true for a significant fraction of satellites.

It should be noted that Paper I did not change the treatment of tidal
stripping of hot gas, so that large amounts of hot gas are still
removed from satellites even in low-mass groups. The fact that
satellites do not accrete from the IGM and that their stars and cold
gas can be totally disrupted by tidal forces also strongly affects
their star formation efficiency. As will become clear later in this
paper, even without ram-pressure stripping, our treatment of
environmental processes still appears to cause too much quenching of
low mass satellites near the centres of low mass groups.  This only
affects a small fraction of the population, however, since most
satellites are relatively far from the group centre.

The probability that a galaxy is a satellite increases with
environment density.  This induces a correlation between quenching and
environment density. The details of such dependences are affected 
by the specific scalings we assume for environmental
effects. In particular, we assume that ram-pressure stripping happens when:
\begin{equation} \label{eq:ram_pressure}
 \rho_{\rm{sat}}\left(R_{\rm{r.p.}}\right)<\frac{\rho_{\rm{par}}\left(R\right)V_{\rm{orbit}}^2}{V_{\rm{sat}}^2},
\end{equation}
where $\rho_{\rm{sat}}(R_{\rm{r.p.}})$ is the hot gas density of the
satellite at radius $R_{\rm{r.p.}}$ from its centre, $V_{\rm{sat}}$ is
the virial velocity of the subhalo at infall, $\rho_{\rm{par}}(R)$ is
the hot gas density of the parent dark matter halo at the current
radius $R$ of the satellite, and $V_{\rm{orbit}}$ is the orbital
velocity of the satellite. Hot gas is assumed to be tidally stripped
at the same rate as dark matter (which is treated fully consistently
in the original simulation) while stars and cold gas are stripped when
the total mass density of the satellite within its half-light radius
falls below the density of the halo within the pericentre of the
satellite's orbit:
\begin{equation} \label{eq:tidal_strip}
 \rho_{\rm{sat}}(R_{{\rm sat},1/2})<\rho_{\rm{DM,halo}}(R_{\rm peri}).
\end{equation}

In combination, these various forms of stripping result in more
efficient quenching in denser parent dark matter halos, closer to
their centres, for larger orbital velocities and for lower satellite
masses.

\begin{figure*}
\centering
\includegraphics[width=17.9cm]{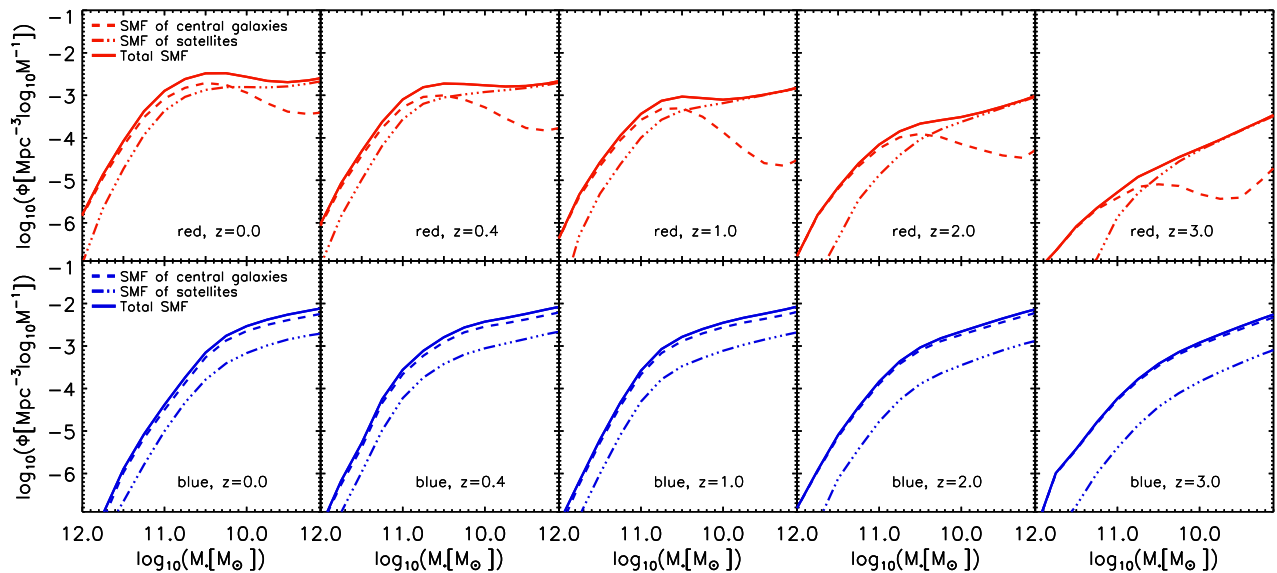}
\caption{Evolution of the stellar mass function for all (solid lines),
  central (dashed lines) and satellite galaxies (dotted-dashed lines).
  Predictions for red and blue galaxies are shown in the top and
  bottom panels, respectively.}
\label{fig:smf_satellites}
\end{figure*}

\subsection{Separation between star-forming/blue and quenched/red galaxies.}
\label{subsec:separation}
In this paper we will use varying cuts to
separate star-forming/blue from quenched/red galaxies in our model. Our approach
is to use the same observables when selecting in the models
as used in the observational studies we compare with. We do, however,
adjust the dividing line so that it lies exactly at the bottom of the valley separating
the two populations in the model. This will allow us to directly compare the relative
numbers of galaxies in each population in the model and in the observed samples
independently of any differences on the global colour distributions arising from 
systematics in the stellar population synthesis or dust modelling.

In the next subsection and
in subsection~\ref{subsec:schechter}, where stellar mass functions and the corresponding 
parameters resulting from Schechter function fits are presented, we will use the same
distinction between red and blue objects as in Paper I. This corresponds to a cut
at $u-r=1.85-0.075\times \rm{tanh}\,((M_r+18.07)/1.09)$, at $z=0$, and in the $UVJ$ colour
plane at higher redshifts. The division between populations is shown in Fig.~4 of Paper I and corresponds to
$U-V=1.3$ and $(U-V) = (V-J) \times a(z) + b(z)$, where $a(z)=[0.28, 0.30, 0.32, 0.38]$ 
and $b(z)=[1.21,1.18,0.99,0.79]$, respectively at $z=0.4$, 1.0, 2.0 and 3.0.

Throughout sections~\ref{sec:agn} and~\ref{sec:env}  
(except for subsection~\ref{subsec:schechter} mentioned above)
passive and star-forming galaxies are separated in the $U-B$ versus stellar mass 
plane, when comparing with \citet{Peng2010} data, and using their SSFR, 
when comparing with \citet{Wetzel2012} data. In detail, red galaxies are defined to have
$U-B>\log_{10}(M_{\star}[\Msun]) \times a(z) + b(z)$, where $a(z)=[0.059, 0.041, 0.052]$ 
and $b(z)=[0.49,0.62,0.53]$ (respectively at $z=0.05$, 0.40 and 0.86), and
$\log_{10}($SSFR[yr$^{-1}])<-11.0$. When comparing model results to the clustering data 
as a function of colour from \citet{Guo2010}, in section~\ref{subsec:correlation_bycolor}, we
select red galaxies using $g-r>\log_{10}(M_{\star} [h^{-2}\Msun ])\times 0.054+0.11$.
Finally when comparing to observations of one-halo conformity from \citet{WangWhite2012}, in 
section~\ref{sec:conformity}, we use $(g-r)^{0.1} >  \log_{10}(M_{\star}[\Msun])\times 0.09 -0.18$.

\subsection{The impact of quenching on different galaxy types -
  stellar mass functions for red and blue satellite and central
  galaxies.}
\label{subsec:smf_satellites}

The two types of quenching process affect different types of galaxies
differently. This can be seen in Fig.~\ref{fig:smf_satellites} which
shows the evolution of the stellar mass functions for satellites
(dotted-dashed lines), central galaxies (dashed lines) and the two
together (solid lines). Predictions are given for passive (red)
galaxies in the upper panels and for star-forming (blue) galaxies in
the lower panels (the separation between red and blue galaxies is described in the 
previous subsection). At all the redshifts considered, star-forming
galaxies are predominantly centrals (the dashed lines are almost on
top of the solid lines in the bottom panels). Massive red galaxies are
also mostly centrals while low-mass passive galaxies are primarily
satellites (top panels). Clearly, low-mass galaxies are quenched
predominantly by environmental processes, and high-mass galaxies by
AGN effects. The radio mode mechanism adopted in our model assumes
feedback from hot-mode accretion to depend on the product of the black
hole and hot gas masses, effectively setting a constant stellar mass
threshold for quenching. This causes model galaxies to quench at the
observed rate, at least since $z\sim 2$, and the characteristic
stellar mass $M_{\star}$ of star-forming objects to remain
approximately constant, also as observed.

In the following sections we will compare our model in detail with
observations of quenching as a function of stellar mass and of a
variety of measures of environment. We note that for most measures,
AGN and environmental affects do not separate cleanly. This is a
consequence of the strong coupling between different aspects of galaxy
formation. For example, although AGN feedback is an internal process
affecting massive galaxies, the growth in mass both of galaxies and of
their central black holes is a strong function of environment, leading
to an enhancement of AGN quenching in high density regions, perhaps
explaining the impact of environment on the quenching of central
galaxies recently noticed by \citet{Knobel2015}. Such correlations may
also be responsible, at least in part, for the observed
conformity between central and satellite galaxy properties
\citep{Weinmann2006b, Kauffmann2010, Wang2012}. This we will analyse
in Sec~\ref{sec:conformity}.

\section{Radio mode quenching of massive galaxies}
\label{sec:agn}
The joint dependence of the fraction of quenched galaxies on stellar
mass and on environment density, as characterised by halo mass,
halocentric distance, nearest neighbour distance or smoothed galaxy
density, was first quantitatively established using SDSS data
\citep{Kauffmann2004, Li2006, Baldry2006, Yang2009} and was then
extended to higher redshift by \citet{Peng2010,
  Peng2012}. Observationally, this dependence seems to separate
relatively well, with stellar mass weakly influencing quenching as a
function of environment and vice-versa. Separability of this kind
arises naturally in semi-analytic models as a result of their
representation of environmental and AGN feedback processes. Since our
new model modifies the treatment of these processes in an attempt to
achieve better agreement with observation, detailed and quantitative
comparison with the real data is a crucial test of its success. In
this section we focus on trends with stellar mass that can test our
new AGN feedback implementation. In sections~\ref{sec:env} and
\ref{sec:clustering} we focus on the impact of quenching as a function
of environment and in section~\ref{sec:conformity} we will look at
whether the combination of quenching effects can lead to the observed
conformity between central and satellite properties.

\begin{figure}
\centering
\includegraphics[width=8.0cm]{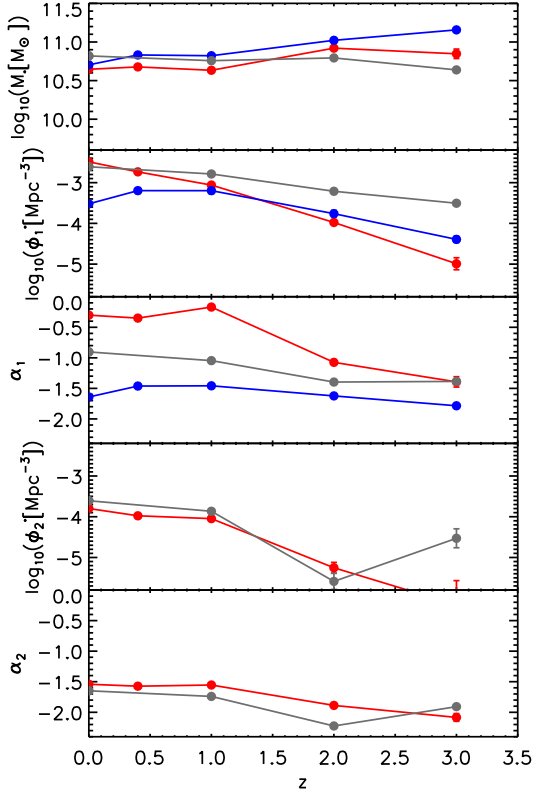}
\caption{The evolution of the parameters of \citet{Schechter1976} fits
  to the stellar mass functions of all (solid grey) passive (solid
  red) and star-forming galaxies (solid blue). Following
  \citet{Peng2010} we fit the simulation data to the sum of two
  Schechter functions (assumed to have the same $M_*$) in the first
  two cases, but use only a single Schechter function for the
  star-forming galaxies. }
\label{fig:schechter}
\end{figure}

\subsection{AGN quenching from the non-evolving stellar mass function
  of blue galaxies} 
  \label{subsec:schechter}
  In Paper I we showed that our new model is
consistent with the observed evolution of the number density of red
and blue galaxies as a function of stellar
mass. Fig.~\ref{fig:schechter} shows the parameters resulting from
Schechter function fits to the evolution of the stellar mass function
of all, of red, and of blue galaxies (the red and blue stellar mass
functions were shown in Fig.~\ref{fig:smf_satellites} as solid
lines). Following \citet{Peng2010}, we have fit the ``red'' and the
``all'' mass functions with double Schechter functions assuming the
same $M_*$ for both components, whereas we fit the ``blue'' mass
function at each time with a single Schechter function 
(the separation between red and blue galaxies is described in subsection~\ref{subsec:separation}).

Not surprisingly, perhaps, given that the evolution of the stellar
mass function and the passive fraction were the primary observational
constraints on our model, the behaviour of the parameters in
Fig.~\ref{fig:schechter} is very similar to that which
\citet{Peng2010} found in the observational data they analysed. The
stellar mass function of star-forming galaxies varies little with
redshift, and indeed, the three parameters combine in such a way that
the actual change in the total number of objects
from $z=0$ to $z=3$ is substantially
less than the change in $\phi^*$ over the full range of stellar masses
shown in Fig.~\ref{fig:smf_satellites}. The star-forming and passive
populations have similar characteristic stellar masses, $M_*$, which
vary little with redshift, and the low-mass slopes of their dominant
components, $\alpha_1$, differ by about unity and have values similar
to those found by \citet{Peng2010} (in the redshift range considered in 
the observational study, $z \le 1$). The normalisation, $\phi^*$,
increases strongly with time for both components of the passive galaxy
stellar mass function, and the low-mass slope of its subdominant
component, $\alpha_2$, is similar to that of star-forming galaxies.
Many (though not all) these features are similar to those of the
simple toy model which \citet{Peng2010} constructed to fit their
observations; the clearest differences are seen at $z \sim 2$, well beyond the
redshift range $0<z\leq 1$ where they concentrated their detailed
analysis.

\begin{figure}
\centering
\includegraphics[width=8.6cm]{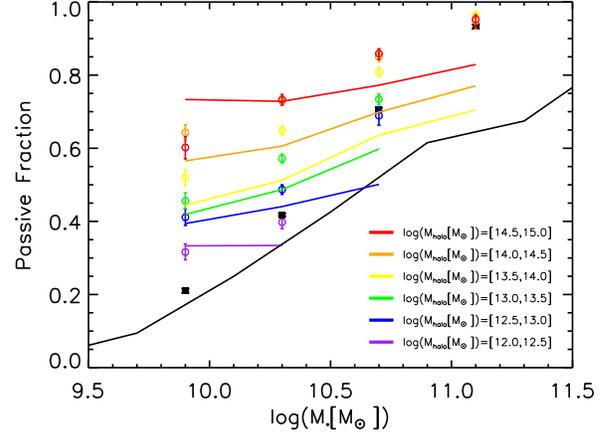}
\caption{The predicted fraction of passive galaxies as a function of
  stellar mass (solid lines) compared with observations compiled by
  \citet{Wetzel2012} (symbols). Black lines and symbols represent the
  relation for central galaxies while coloured lines and symbols are
  for satellites in groups of differing mass. Both the observed and
  the model passive galaxies were selected by specific star formation
  rate ($\log_{10}($SSFR[yr$^{-1}])<-11.0$).}
\label{fig:wetzel_passive_fraction_mass}
\end{figure}

The rate at which massive galaxies quench can be inferred from the
evolution of the massive end of the stellar mass functions.  Since
$M_*$ varies little with redshift and $\alpha_1$ gets shallower at late times
for high-mass, passive
galaxies, the evolution in their $\phi_1^*$, by factors of about 300,
20 and 3 since $z=3, 2$ and 1, respectively, is a lower limit of the impact of
the quenching process. The fact that $M_{\star}$ for blue galaxies is
roughly constant, at least for $z \le 1$, shows that quenching occurs with high
probability once stellar masses become comparable to $M_*$, thereby
terminating galaxy growth through star formation, and that this
probability depends weakly on redshift. As \citet{Peng2010} show,
it is exactly these assumptions in their toy model which enable it
to reproduce the observed phenomenology. They hold approximately in
our physical, $\Lambda$CDM model as a result of the way that our
assumptions about AGN growth and feedback interact. Black holes grow
rapidly at moderate redshift in galaxies of stellar mass $\sim
10^{10}{\rm M_\odot}$ as gas-rich mergers build up their bulges. The
merger products typically have halos of mass $\sim 10^{12}{\rm
  M_\odot}$ in which cooling times become long enough to allow the
build up of a hot gas envelope.  The combination of massive black hole
and hot envelope is then able to generate enough ``radio-mode''
feedback to cut off further cooling and quench the galaxy.  For the
scalings adopted in our model, the transition occurs at a halo mass and
central galaxy stellar mass which are almost independent of redshift.
This will be further investigated in a companion paper (Henriques et al. in prep)
and is similar to the findings of \citet{Dubois2015} and \citet{Bower2016}.

It should be noted that mergers play a relatively small role in the
growth of massive galaxies in this model, and hence in shaping the
high-mass cut-off in the stellar mass functions. These extreme objects
are mostly central galaxies in groups and clusters, where mergers
contribute significant material to the intracluster stars. Nevertheless, 
because the boundary between these two components is ill-defined,
different choices emphasise the importance of mergers differently, and
can lead to apparent differences at high mass both in observational
and in simulated stellar mass functions \citep{Bernardi2013,
  Kravtsov2014, Schaye2015}.

\begin{figure*}
\centering
\includegraphics[width=18.9cm]{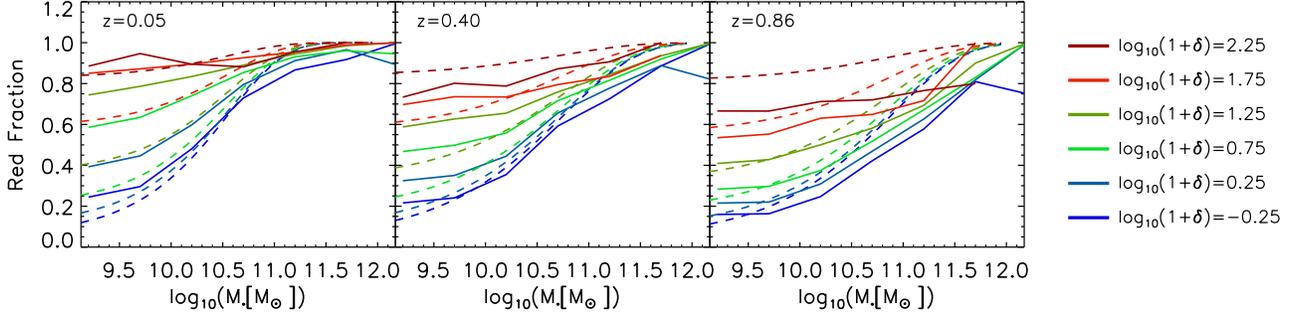}
\caption{Solid lines show the evolution of the fraction of galaxies
predicted to be passive (red) as a function of stellar mass from $z=0.86$ (right)
through $z=0.4$ (centre) down to $z=0.05$ (left). Different colours represent different
  $0.5$~dex ranges in density (as inferred from the distance to the
  5th nearest neighbour) centred on the values indicated by the labels
  to the right of the panels. Observational results from SDSS and
  COSMOS taken from \citet{Peng2010} are shown as dashed lines.}
\label{fig:peng_red_fraction_mass}
\end{figure*}

\subsection{AGN quenching from the fraction of passive/red
  galaxies versus stellar mass}

In this subsection we focus on predictions for the fraction of passive
galaxies as a function of stellar and halo mass. Our analysis differs
from that of red fraction as a function of stellar mass in Paper I, in
that we here compare with observational data with better statistics at $z<1$, we use a
bluer colour to classify galaxies ($U-B$ instead of $u-r$), and we
separate galaxies according to instantaneous star formation rates, in
addition to using colours (the details of the cuts used are presented in subsection~\ref{subsec:separation}).  
These modifications allow us to compare
more closely with the observational analyses of \citet{Peng2010} and
\citet{Wetzel2012}.

\subsubsection{AGN quenching from the fraction of passive galaxies
  versus stellar mass at $z\approx0.1$}

Fig.~\ref{fig:wetzel_passive_fraction_mass} shows predictions for the
fraction of passive galaxies as a function of stellar mass at
$z\approx0.1$ (solid lines) and compares them with SDSS data compiled
by \citet{Wetzel2012}. The colours differentiate results for satellite
galaxies living in groups/clusters of different mass (for the
observations, these masses were estimated by an abundance matching
argument rather than measured directly). Black lines and symbols give
corresponding results for central galaxies.  In the observational
samples, the quenched fraction increases both with stellar mass and
with halo mass. The latter dependence (together with the fact that
satellites of a given stellar mass are more likely to be quenched than
centrals) is an environmental effect that we will discuss in more
detail in section~\ref{sec:env}. Our simulation reproduces these
trends quite well although with some quantitative differences. In
particular, the increase in passive fraction with stellar mass for
central galaxies is less steep in the simulation than is observed.
Quenching is also somewhat more frequent than observed at low stellar
mass and is slightly less complete than observed at high stellar mass.

These discrepancies are larger than found in Paper I.  As we confirm
in the next subsection, this seems in part to result from using
directly estimated SSFRs rather than
colours to separate passive from star-forming galaxies. They are also
more noticeable in the $U-B$ colours used below than in the $u-r$
colour used in Paper I. This may indicate that although more massive
galaxies are dominated by older populations in our model, as
observationally required, too many of them still have residual star
formation at late times (the same conclusion was recently reached by \citealt{Luo2016}).  
Although we have increased the efficiency of
radio-mode feedback at later times, this is not enough to suppress
star formation completely. Massive objects that 
remain star forming at late times are characterised by relatively "quiet" 
accretion histories, with no major merger below $z=2$ (these represent $\sim20\%$ of the
population at $\log_{10} (M_*[\Msun]) = 11.0$). This leads to 
black hole masses below average and little AGN activity and seems 
to indicate that our model for black hole growth excessively relies on
mergers. We checked that the problem is alleviated if we allow black hole 
growth also in disk instabilities and  we will include this modification in future 
versions of our model.

\subsubsection{Evolution of the red fraction as a function of stellar mass}
\label{subsec:peng_redfract_stellarmass}

In this subsection we compare our simulation directly to the data
which \citet{Peng2010} used to quantify quenching over the redshift
range $0<z\leq 1$. They separated passive from star-forming galaxies
on the basis of rest-frame $U-B$ colour, and then used large
observational samples from SDSS and COSMOS to estimate the evolution
of quenched fraction as a function of stellar mass. 
Fig.~\ref{fig:peng_red_fraction_mass} compares our current model
(solid lines) to the fits which \citet{Peng2010} made to their
compilation of observations (dashed lines). The different panels refer
to different redshifts ($z=0.05$, 0.40 and 0.86) while the different
lines within each panel split the galaxies by environment
density (the separation between red and blue galaxies is described in subsection~\ref{subsec:separation}). 
The proxy for
environment density used in the simulation is directly analogous to
that used by \citet{Peng2010} and is based on the projected distance
to the 5th nearest neighbour with a redshift difference (peculiar
velocity + Hubble flow) less than $1000\,\rm{km\;s^{-1}}$. Such
distances are calculated for all galaxies with
$\log_{10} (M_*[\Msun]) \geq 9.0$ but only bright neighbours
($M_{B,\;\rm{AB}}\le-19.3-z$ at $z\le0.7$ and
$M_{B,\;\rm{AB}}\le-20.5-z$ at $z>0.7$) are counted when estimating
environment density. The density itself is given by
$\Sigma_{5,i}=5/\pi d^2_{5,i}$ and the normalised density used in the
plots by $1+\delta_{5,i}=\Sigma_{5,i}/\Sigma_{5,m}$ where
$\Sigma_{5,m}$ is the mean density of ``bright'' galaxies at given
redshift.

Model and observations agree reasonably well in the slope of the
relations and in their density dependence at all redshifts. In
particular, there is a substantial improvement at $z\sim0$ compared to
early attempts at modelling such as that of \citet{Baldry2006}. The most 
obvious discrepancies occur at high mass and at high overdensity, for $z>0.05$, where 
quenching is less effective than observed (at $z=0.05$ only the discrepancy at high mass remains). 
As explained in the previous section,
this discrepancy is evident in terms of specific star formation, it is
still noticeable for $U-B$ colours, but it is absent for red galaxies in
terms of $u-r$. This indicates that the ages in the model and in the
observations are comparable but there is too much ongoing star formation at 
high mass and environment density in the model, particularly at $z\sim 1$.

While the slopes tell us about AGN/stellar mass quenching efficiency,
the different colours indicate the impact of environment. In this case
it seems that the model matches observations at the higher redshifts,
but over-estimates the red fraction in high density environments at
later times (compare solid and dashed lines for each
colour at low masses which are not affected by insufficient 
quenching by AGN feedback). The discrepancy may result from a mismatch in the density
calculation or from overly high impact of the environment in the
model. This will be analysed in detail in the section~\ref{sec:env}.

\begin{figure}
\centering
\includegraphics[width=8.6cm]{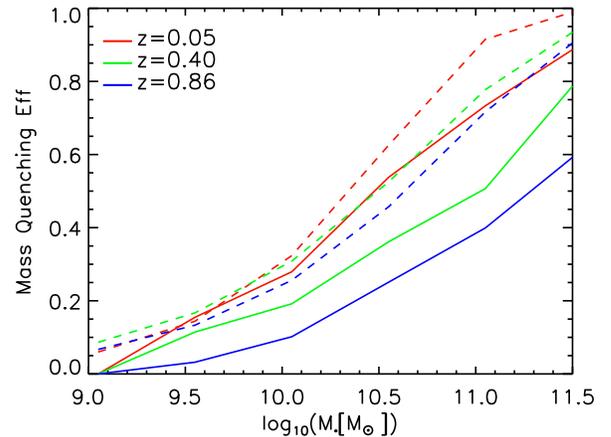}
\caption{Evolution of the mass-quenching efficiency from $z=0.86$ to
  0.05. Predictions from our simulation are shown as solid lines, while
  the fits which \citet{Peng2010} used to represent their compilation
  of observational data are shown as dashed lines.}
\label{fig:peng_mass_quench_evo}
\end{figure}

\begin{figure*}
\centering
\includegraphics[width=17.9cm]{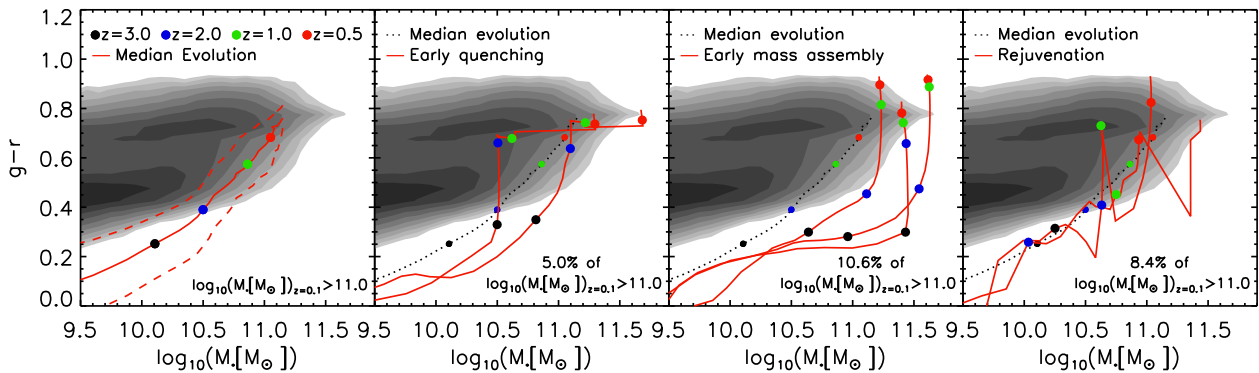}
\caption{Evolutionary tracks of galaxies with $\log_{10} (M_*[\Msun]) \ge 11.0$ at 
$z=0.1$ in the $g-r$  versus stellar mass plane (solid red lines). In the left panel the 
median and 16th and 84th percentiles are shown for the evolution of all galaxies 
(repeated as a dotted black line in the other panels) while the evolution of individual 
galaxies is plotted in other panels. Various redshifts are marked with filled circles of 
differing colour.}
\label{fig:mass_color}
\end{figure*}

\subsection{Evolution of the mass-quenching efficiency}
\label{subsec:agn_quench_eff_evo}

\citet{Baldry2006} and \citet{Peng2010} showed that the mass and
environment dependences of the probability that a given galaxy is red
(passive) appear to be separable, meaning that the probability can be
written as a product of two factors, one depending only on stellar
mass and the other depending only on environment density. Analysis in
terms of these so-called quenching efficiencies allows us to focus on
the impact of AGN feedback and environment independently.  Here we
compare the observationally inferred evolution of the mass-dependent
efficiency to the predictions of our model. Following \citet{Peng2010}
the mass-quenching efficiency can be defined as:
\begin{equation} \label{eq:mass_quench} 
\varepsilon_m(m,m_0,\rho ,z) =
  \frac{f_{\rm{red}}(m,\rho ,z)-f_{\rm{red}}(m_0,\rho ,z)}{1 -f_{\rm{red}}(m_0,\rho ,z)},
\end{equation}
which gives the excess red fraction at stellar mass $m$ relative to
the red fraction at the lowest mass considered $m_0$ for galaxies in
environments of density $\rho$ at redshift $z$. Both observationally
and in our model, this excess is approximately independent of $\rho$,
so we will present results averaged over environment density.

Fig.~\ref{fig:peng_mass_quench_evo} compares observational and
simulation results for the evolution of environment-averaged
mass-quenching efficiency. Solid lines show our model predictions
while dashed lines show fits to observational data from COSMOS and
SDSS by \citet{Peng2010}. Different colours indicate different
redshifts. The trends highlighted in the previous subsection are
clearly visible in this plot. The mass-quenching efficiency increases
with stellar mass and decreases with redshift over the full range of
mass and redshift shown. Overall, the slopes of the model relations
are roughly consistent with those observed, indicating that our
revised AGN feedback implementation reproduces the observed evolution
of the mass-quenching efficiency quite well. In the model, the trends
with redshift and stellar mass result from the close but
redshift-dependent relations between stellar mass and the properties
that control hot gas accretion onto black holes, namely black hole
mass and halo hot gas mass. As previously highlighted, a discrepancy
remains for the highest masses where AGN feedback is not efficient
enough in our model.

\subsection{Individual galaxy tracks through the
  colour-stellar mass diagram}

As described above, various quenching mechanisms can move galaxies
from the star-forming (blue) to the passive (red) population.  For
example, galaxies may develop a large enough black hole for AGN feedback
to prevent cooling of hot gas to replenish their cold interstellar
medium, or this medium may be removed by interaction with their
environment. Once quenched, ongoing black-hole feedback keeps the
majority of central galaxies quenched in our models, while further
stripping keeps satellites quenched until they merge with the central
object of their group/cluster. Nevertheless, it is not uncommon for
quenched central galaxies to become star-forming again after major
episodes of gas accretion, either from the diffuse intergalactic
medium or from merging. AGN feedback is most effective in massive
galaxies, making them predominantly red by $z=0$. It has little effect
at low-mass, so that dwarf galaxies remain blue unless they are
strongly affected by environmental processes\footnote{Only $\sim40\%$
  of galaxies are satellites at $z=0$ and only a fraction of those
  will be quenched.}.

It is particularly informative to plot the evolution of massive
galaxies in a colour-stellar mass diagram, where quenching takes them
from being predominantly star-forming at $z>2$ to predominantly
passive at $z=0$. The stochasticity of black hole and hot gas growth
means that AGN feedback quenches different galaxies at different times
leading to significantly different tracks in this
diagram. Fig.~\ref{fig:mass_color} illustrates this; red lines
represent the time evolution of individual massive galaxies in the
$(g-r)$-stellar mass plane ($\log_{10} (M_*[\Msun]) \ge 11.0$ at $z=0.1$). 
In the left panel the median and 16th and 84th percentiles are shown for 
 the evolution of all galaxies (repeated as a dotted black line in the other panels) 
 while the evolution of individual galaxies is plotted in other panels.
Coloured dots along these lines mark
specific redshifts; black for $z=3$, blue for $z=2$, green for $z=1$
and red for $z=0.5$. The end of the line corresponds to
$z=0.1$. Over-plotted in grey contours is the full galaxy distribution
at $z=0.1$. When comparing the tracks with this distribution it should
be remembered that the corresponding distributions at earlier times
were significantly bluer. As described in
section~\ref{subsec:radio_quench}, properties in addition to stellar
mass contribute to the impact of AGN feedback on massive galaxies and
hence their evolution in this figure. \citet{Terrazas2016a} carries out
a detailed analysis of this issue in the context of our model for the
particular case of central galaxies in Milky-Way mass halos. 

The left panel of Fig.~\ref{fig:mass_color} shows the median evolutionary
track for massive galaxies. These are star-forming at early times, but quench 
once their central black hole becomes sufficiently massive. The transition 
typically happens between redshifts 2 and 1 (the blue and green dots, respectively).
Galaxies travel gradually from the blue side to the red side of the
star-forming population as they grow in mass, then move rapidly to the
passive population once feedback is able to prevent cooling and
accretion of new material and the remaining cold gas is
exhausted (seen by the increased steepness of the track below $z=2$, 
blue symbol). This results in a broad star-forming sequence and a paucity
of galaxies in the ``green valley'' between the two populations. Once
massive galaxies reach the red sequence, the absence of further
star-formation and also of minor mergers once they become satellites
in groups or clusters causes most systems to grow rather little in
mass. It should be noted that the comparative rarity of massive blue
galaxies at $z=0$ is perfectly compatible with this path. There were
more such galaxies at $z=2$, but most have now become passive, a
sequence often referred to as ``down-sizing'' of star-formation
activity.

The left middle panel of Fig.~\ref{fig:mass_color} shows a less
common, but still significant, evolutionary pathway in which galaxies
quench early and then grow substantially through mergers: galaxies 
which become red at $z \ge 2$ ($5.0\%$ of all galaxies with $\log_{10} (M_{*(z=0.1)}[\Msun]) \ge 11.0$ 
and $17.0\%$ of all galaxies with $\log_{10} (M_{*(z=0.1)}[\Msun]) \ge 11.5$).  
Such massive galaxies begin life as the central objects of high-mass halos that
undergo early gas-rich mergers, resulting in substantial black hole
growth. This can produce enough AGN feedback at this early stage to
quench star formation entirely. Later growth then occurs through
accretion of satellite galaxies onto the central object of the
group/cluster. Another extreme case, this time of early mass-assembly,
is shown in the middle right panel: galaxies which have increased their mass by 
less than a factor of 2 since $z=2$ ($10.6\%$ of all galaxies with $\log_{10} (M_{*(z=0.1)}[\Msun]) \ge 11.0$ 
and $21.6\%$ of all galaxies with $\log_{10} (M_{*(z=0.1)}[\Msun]) \ge 11.5$). One of these galaxies is 
already as massive as $10^{11.5}\Msun$ at $z=3$ (right most black dot). While such early
growth of massive objects may seem ``anti-hierarchical'', it is
clearly fully compatible with $\Lambda \rm{CDM}$ structure formation
once baryonic physics are taken into account. Finally, the right panel
of Fig.~\ref{fig:mass_color} shows examples of galaxies that return to
the blue population after being quenched: galaxies which have decreased
their $(g-r)$ colour by more than 0.1 after reaching the red population ($(g-r)$=0.6)
 ($8.4\%$ of all galaxies with $\log_{10} (M_{*(z=0.1)}[\Msun]) \ge 11.0$ 
and $10.2\%$ of all galaxies with $\log_{10} (M_{*(z=0.1)}[\Msun]) \ge 11.5$). This is normally a
consequence of direct accretion of cold gas from satellites. Since
such galaxies contain a massive black hole that previously shut down
cooling, the return to the passive population tends to be quick.

\begin{figure*}
\centering
\includegraphics[width=17.9cm]{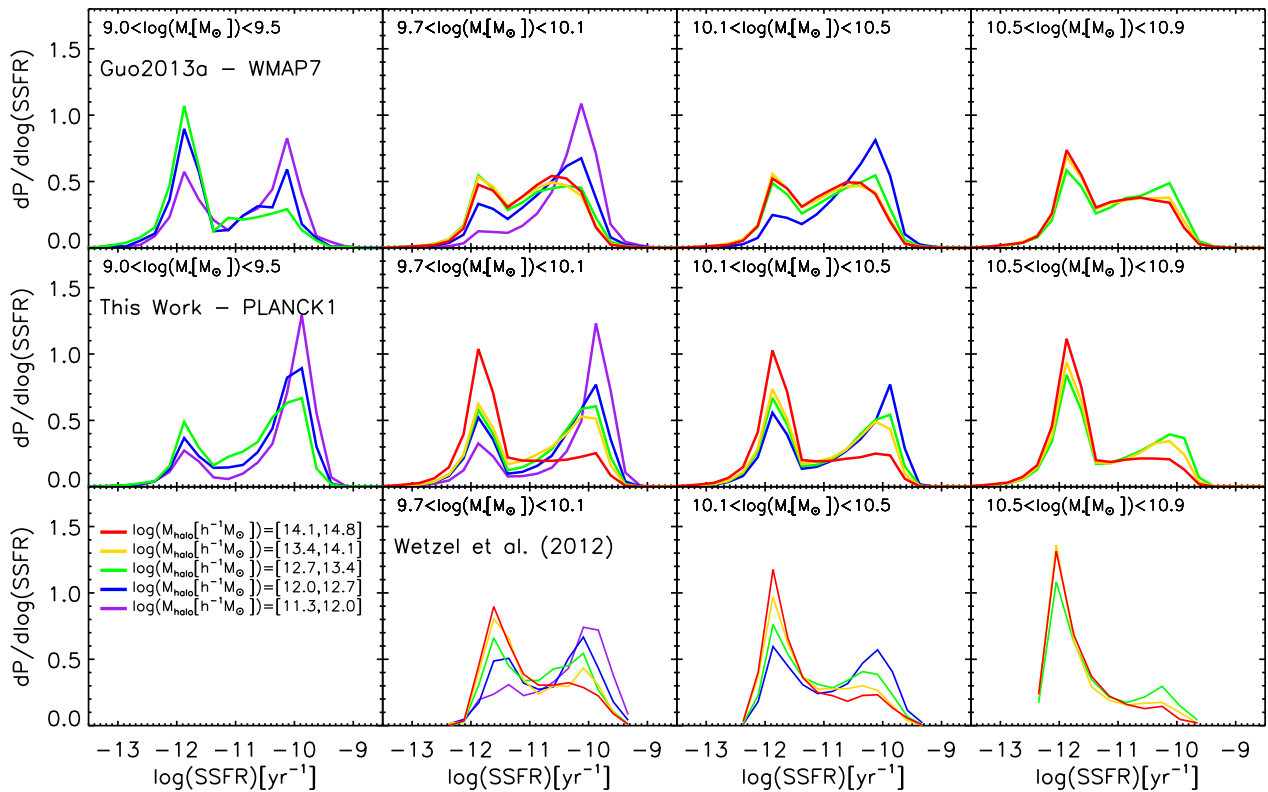}
\caption{SSFR histograms for different halo masses (different colors)
  and for different stellar mass bins (low stellar mass left panels,
  high stellar mass right panels). The top row shows predictions from
  \citet{Guo2013}, the middle row shows predictions from the model of
  this paper, while the bottom row shows observational results
  compiled by \citet{Wetzel2012}.}
\label{fig:wetzel_ssfr_halomass}
\end{figure*}

\section{Environmental quenching}
\label{sec:env}

In this and subsequent sections we will focus on environmental
quenching, an effect caused in our model by the removal of gas from
satellite galaxies as they orbit within the halo of a larger
system. As in the previous section, we will consider the fraction of
galaxies of given stellar mass which are passive or red, but now as a
function of environment rather than of stellar mass. In the current
section, we will study quenched fractions as a function of host halo
mass, of distance to group centre, and of local number density of
galaxies. In section~\ref{sec:clustering} we will study how
environmental quenching affects the mass- and colour-dependent
autocorrelation functions of galaxies. Finally, in
section~\ref{sec:conformity} we will look at how the combined effects
of AGN and environment result in the observed conformity between the
properties of central and satellite galaxies. We emphasise that
despite ram-pressure stripping being switched off in low-mass groups
in our current model, satellites still experience strong environmental
effects as a result both of the suppression of primordial infall and 
of tidal stripping by the group potential.

\subsection{Passive fraction versus halo mass}
\label{sec:wetzel_passivefrac_halomass}

Fig.~\ref{fig:wetzel_ssfr_halomass} shows specific star formation rate
distributions for satellite galaxies in various bins of stellar mass
and group/cluster mass. From left to right, the panels give results
for galaxies of increasing stellar mass, as indicated by the legend.
Within each panel, different coloured lines correspond to different
group masses as indicated by the key at bottom left. The top row shows
results for the model of \citet{Guo2013}, the middle row for the model
of this paper, and the bottom row for SDSS data as compiled by
\citet{Wetzel2012}. All distributions have been normalised to enclose
the same area. In the models, the halo mass is obtained directly from
the dark matter simulation, after scaling to the appropriate
cosmology. For the SDSS data, an abundance matching approach was taken
in order to assign sets of central+satellite galaxies to haloes of
differing mass. As explained in Paper I, in order to match the observed
distribution of unclassified galaxies with no emission lines, that
were assigned a value of SFR based on SED fitting, model galaxies with
$\log_{10}($SSFR[yr$^{-1}])<-12$ have been assigned a random Gaussian
value centered at $\log_{10}($SSFR[yr$^{-1}])=-0.3\log_{10}(M_*[\Msun])-8.6$ and with dispersion
0.5.

Some clear trends can be identified in the observational data (bottom
row of Fig.~\ref{fig:wetzel_ssfr_halomass}). Massive galaxies are
predominantly passive, irrespective of the mass of the group they
reside in (lines of all colours peak at low SSFR in the bottom
right-most panel). Conversely, in the high-mass clusters, galaxies are
predominantly passive, irrespective of their stellar mass (the red
lines peak at low SSFR in all three panels of the bottom row). For
intermediate mass galaxies, the passive fraction increases strongly
with increasing halo mass (middle panels of the bottom row).

\begin{figure}
\centering
\includegraphics[width=8.6cm]{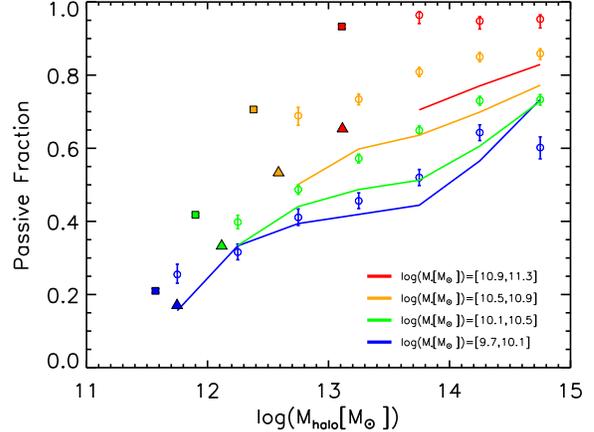}
\caption{The fraction of satellite galaxies that are passive as a
  function of host halo mass (defined as those with $\log_{10}($SSFR[yr$^{-1}])<-11$). 
  Solid lines and open circles represent
  model predictions and observational data from SDSS (again from
  \citealt{Wetzel2012}), respectively. Different colours refer to
  galaxies of different stellar mass as indicated in the legend.
  Filled triangles and filled squares indicate the passive fraction and
  mean halo mass of {\it central} galaxies with the corresponding
  stellar mass in our models and in the SDSS respectively }
\label{fig:wetzel_passivefraction_halomass}
\end{figure}

These trends are reproduced qualitatively by our model (the middle row
in Fig.~\ref{fig:wetzel_ssfr_halomass}). Galaxies in higher mass
groups/clusters, even those with lower stellar and black hole masses,
suffer stronger environmental effects and are more likely to be
quenched. This trend extends down to low-mass groups, where
ram-pressure stripping is switched off, because other environmental
processes are still active. In fact, environmental effects are
stronger in our new model than in \citet{Guo2013}, as can be seen from
the larger passive peaks (at $\log_{10}($SSFR[yr$^{-1}])\sim -12$) for
galaxies with $\log_{10} (M_*[\Msun]) > 9.4$ (compare the top and middle
rows in the three rightmost panels). This is caused by the more
efficient supernova feedback introduced by \citet{Henriques2013} which
suppresses the early formation of low-mass galaxies, balancing it with
re-incorporation and primordial infall at later times. Since none of
the replenishing channels are available for satellites, the stronger
feedback results in quicker depletion of gas in such objects. The
impact of the weaker ram-pressure stripping and the lower threshold
for star formation is only evident for the lowest mass galaxies,
resulting in a smaller passive peak (left panel, for
$8.7<\log_{10} (M_*[\Msun]) <9.2$).

A simpler version of the same data is shown in
Fig.~\ref{fig:wetzel_passivefraction_halomass}, where the passive
fraction of satellite galaxies (defined as those with $\log_{10}($SSFR[yr$^{-1}])< -11$) 
is plotted as a function of group/cluster
mass, with different colours representing different stellar mass
ranges. The circles with error bars are SDSS data taken from
\citet{Wetzel2012} while the solid lines are predictions from our
model. Filled triangles and squares mark the passive fraction and mean
halo mass of {\it central} galaxies with the corresponding stellar
mass in our model and in the SDSS, respectively. As seen in the
previous figure, our model is roughly consistent with observation.
The variation in passive fraction with host halo mass (i.e. the slope
of the relation) agrees well, but there is a discrepancy in
normalisation at high stellar mass which, as discussed in previous
sections, reflects the fact that, even with our more efficient AGN
feedback, massive galaxies still seem to have too much ongoing star
formation at late times. This is also seen clearly in the
offset between model and observation in the passive fraction of central
galaxies.

\subsection{Passive fraction versus distance to group centre}
\label{sec:wetzel_passivefraction_radius}

A different test of our implementation of environmental effects comes
from looking at how the fraction of satellite galaxies that are
passive varies with distance from group centre. Although individual
orbits can span a wide range of radii, satellites seen closer to group
centre have, on average, spent more time as group members and have
spent that time in denser regions.  Thus, they are more likely to be
environmentally quenched, and the radial dependence of passive
satellite fractions should be sensitive both to the variation of the
strength of environmental effects with radius and to the timescale on
which star formation is suppressed.

\begin{figure}
\centering
\includegraphics[width=8.6cm]{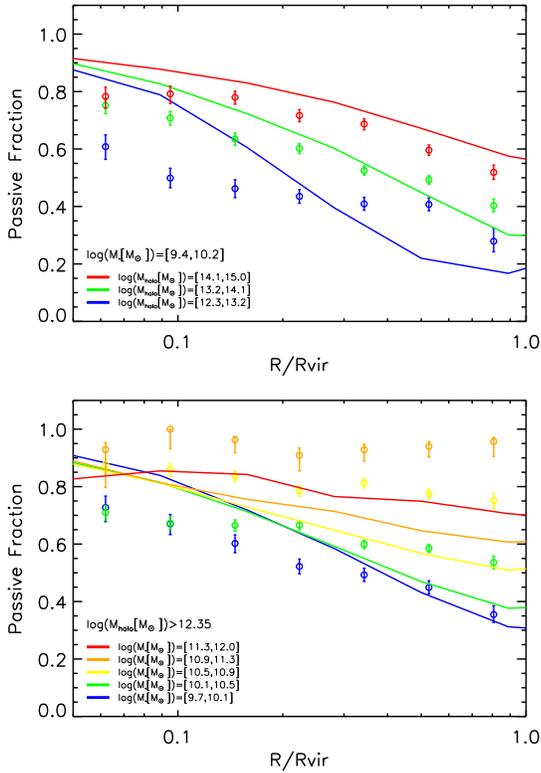}
\caption{The fraction of passive galaxies (defined as those with
  $\log_{10}($SSFR[yr$^{-1}]) < -11$) as a function of
  projected distance from the central galaxy in bins of group mass
  (top) and (satellite) stellar mass (bottom). Model predictions are
  shown as colored lines and SDSS data from \citet{Wetzel2012} as
  coloured symbols.}
\label{fig:wetzel_passivefraction_radius}
\end{figure}

\begin{figure*}
\centering
\includegraphics[width=18.9cm]{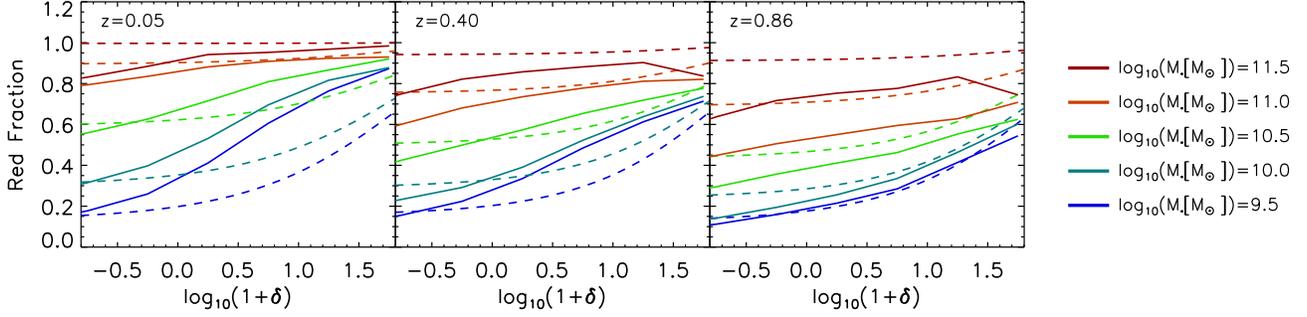}
\caption{The predicted evolution of the fraction of red galaxies as a
  function of local galaxy density as measured by the distance to the
  5th nearest neighbour (solid lines) from $z=0.86$ (right panel), to
  $z=0.40$ (middle panel) to $z=0.05$ (left panel). The different
  colours represent different bins of stellar mass as indicated. Fits
  by \citet{Peng2010} to their compilation of SDSS and COSMOS
  observations are shown as dashed lines.}
\label{fig:peng_redfraction_env}
\end{figure*}

Fig.~\ref{fig:wetzel_passivefraction_radius} shows the fraction of
passive galaxies as a function of projected radius for different
ranges of halo/group mass (top panel) and also for different ranges of
galactic stellar mass (bottom panel). In the first case, all curves
refer to galaxies with $9.4 \leq \log_{10} (M_*[\Msun]) \leq 10.2$,
while in the second, all curves are for galaxies in groups with
$\log_{10} (M_{\rm vir}[\Msun]) > 12.35$. Over the range $0.3 <
R/R_{\rm{vir}} <1$, the model agrees well with the SDSS data compiled
by \citet{Wetzel2012} (top panel and low stellar masses in the bottom panel).  
This is consistent with results shown in
previous sections where similar agreement was found for the population
of satellites as a whole in different stellar masses and halo mass
bins. Nevertheless, it seems clear from both panels that the model
relations are steeper than observed, resulting in too many passive
galaxies near group centre. This is particularly clear for low-mass
satellites and in low-mass groups. We emphasise that this is a
subdominant population, since most satellites orbit at large radii.
This explains why such discrepancies were not evident in the global
trends. In addition, it is once again clear in the bottom panel
that massive galaxies have too much ongoing star-formation.
This is seen at all radii and reflects the inability of our AGN feedback
model to quench all of these galaxies in the field.

Even though, in our current model, ram-pressure stripping acts only in
groups with $\log_{10} (M_{\rm{vir}}[\Msun]) > 14.0$, the impact of environment
still appears too strong in low-mass systems. It seems that the
assumed lack of primordial infall or other gaseous accretion onto
satellites combines with the expulsion of ISM by supernova feedback
and the removal of hot gas haloes by tidal stripping to produce
quenching on too short a time-scale. Despite the significant modelling
improvements introduced recently to address problems pointed out by
\cite{Font2008}, \cite{Weinmann2010}) and \cite{Guo2011} among others,
these remaining discrepancies show that further modifications are
needed. A possibility might be to include interaction-induced star
formation as suggested by \citet{Wang2014}. We have carried out some
checks to see if the problem can be alleviated by relaxing the crude
instantaneous recycling approximation we use to describe stellar
evolution and the attendant return of gas and metals to the ISM. The
mass return can extend star formation in satellites by up to $\sim
2$~Gyr, but this is not enough to solve the problem, since by $z=0$
many of the inner galaxies in our groups and clusters have already
been satellites for 5 or 6 Gyr.

\subsection{Evolution of the relation between red fraction and
 environment density}

In order to compare with the SDSS and COSMOS data compiled by
\citet{Peng2010}, we now use distance to the 5th nearest neighbour
galaxy as an indicator of environment and study how it affects the
quenching of galaxies as indicated by their rest-frame $U-B$
colours (the separation between red and blue galaxies is described in subsection~\ref{subsec:separation}). 
This allows us to test our model against observations up to
$z\sim1$, providing significant constraints on how well it represents
the time evolution of environmental effects.  Details of the density
calculations are described in
section~\ref{subsec:peng_redfract_stellarmass}.

\begin{figure}
\centering
\includegraphics[width=8.6cm]{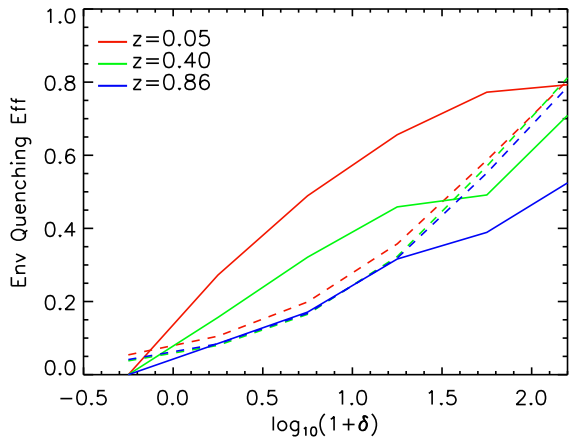}
\caption{Evolution of the environmental quenching efficiency from
  $z=0.86$ to 0.1. Predictions from our model are shown as solid lines
  while those of the fitting function which \citet{Peng2010} use to
  represent their SDSS and COSMOS observations are shown as dashed
  lines.}
\label{fig:peng_envquench_evo}
\end{figure}

Fig.~\ref{fig:peng_redfraction_env} compares our model (solid lines)
to the fits which \citet{Peng2010} used to represent the dependence of
red fraction on environment density which they infer from their
compilation of SDSS and COSMOS data (dashed lines); different colours
show results for different stellar mass ranges. The left, middle and
right panels are for galaxy samples with mean redshift, 0.05, 0.40 and
0.86, respectively. At high redshift (the right panel), the model
agrees quite well with observation, matching the increase in red
fraction from low to high density at all stellar masses 
(compare the slopes of the solid and dashed lines). At later
times, the model relations seem too steep, producing excessive red
fractions at moderate to large over-densities. This is another
reflection of the overly strong quenching in the central regions of
groups and clusters pointed out in the previous section.  It is
interesting that overly strong quenching is only seen at $z<0.5$.

In order to separate environmental from AGN quenching in a
complementary way to that of section~\ref{subsec:agn_quench_eff_evo},
we again follow \cite{Peng2010} and define a redshift-dependent
environmental-quenching efficiency as the excess red galaxy fraction
at each environment density with respect to that at the lowest
density:
\begin{equation} \label{eq:env_quench}
\varepsilon_{\rho}(\rho,\rho_0,m)
= \frac{f_{\rm{red}}(\rho,m)-f_{\rm{red}}(\rho_0,m)}{f_{\rm{blue}}(\rho_0,m)}.
\end{equation}
As explained in section~\ref{subsec:agn_quench_eff_evo}, AGN and
environmental quenching are, to a good approximation, separable,
meaning that the efficiency defined by this equation is at most weakly
dependent on stellar mass. We therefore present results for
$\varepsilon_{\rho}(\rho,\rho_0)$ averaged over stellar mass.

Fig.~\ref{fig:peng_envquench_evo} compares our model's prediction for
this quantity (solid lines) to the analytic fit which \citet{Peng2010}
use to represent their observational data (dashed lines). Blue, green
and red lines represent redshifts $0.86$, 0.40 and 0.05,
respectively. As expected from Fig.~\ref{fig:peng_redfraction_env},
model and observations agree reasonably well at $z=0.86$ 
(except at the highest densities), but at lower
redshifts, the environmental quenching efficiency in the model seems
higher than observed (see also, the end of
section~\ref{sec:wetzel_passivefraction_radius}).  Despite the
considerable improvements achieved with our model updates, this again
indicates that environmental processes remain too strong, particularly
in the inner regions of groups and clusters. With respect to the highest
overdensities (right-most values), it seems that 
our model has the correct environmental quenching efficiency by $z=0.05$ but 
decreases more than observed at higher redshifts. A recent study suggests 
that such evolution may be present at even
higher redshifts ($z>0.9$), with quenching efficiencies 
dropping to 0.2 by $z=1.6$ \citep{Nantais2016}.

\begin{figure*}
\centering
\includegraphics[width=17.9cm]{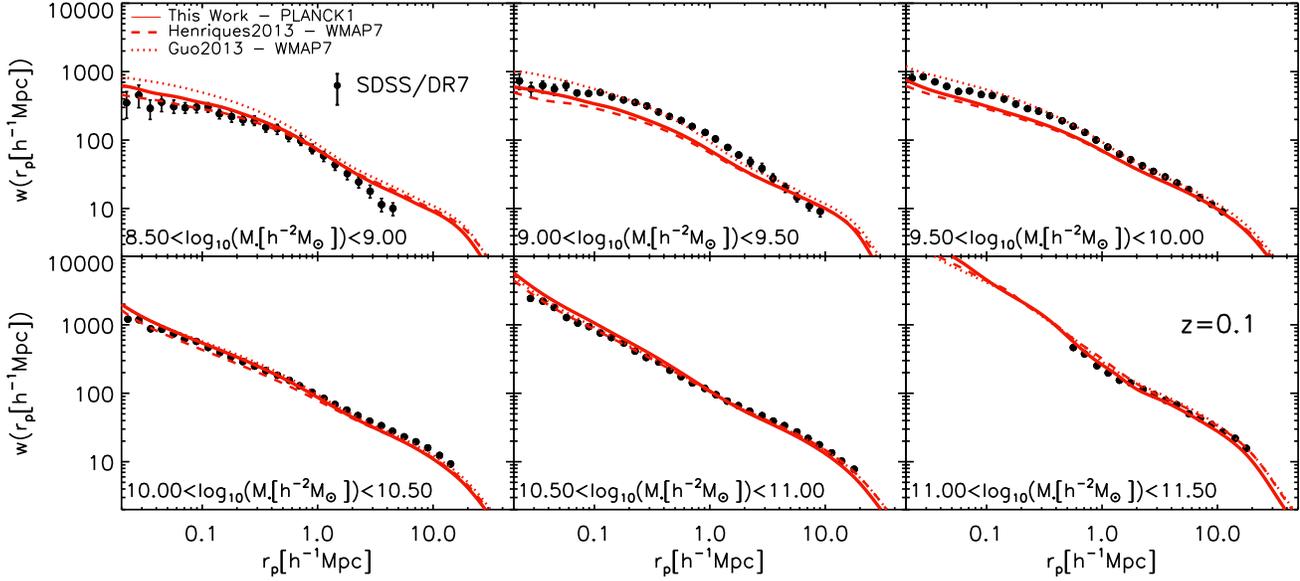}
\caption{Projected 2-point autocorrelation functions for galaxies in
  six disjoint stellar mass ranges at $z=0.1$. Results from our new
  model with its best-fit parameter values (solid red lines) are
  compared with model results from \citet{Henriques2013} (dashed red
  lines) and \citet{Guo2011} (dotted red lines) and observational
  results based on SDSS/DR7 (black symbols, taken from
  \citealt{Guo2011}). }
\label{fig:correlation_z0}
\end{figure*}

\begin{figure*}
\centering
\includegraphics[width=17.9cm]{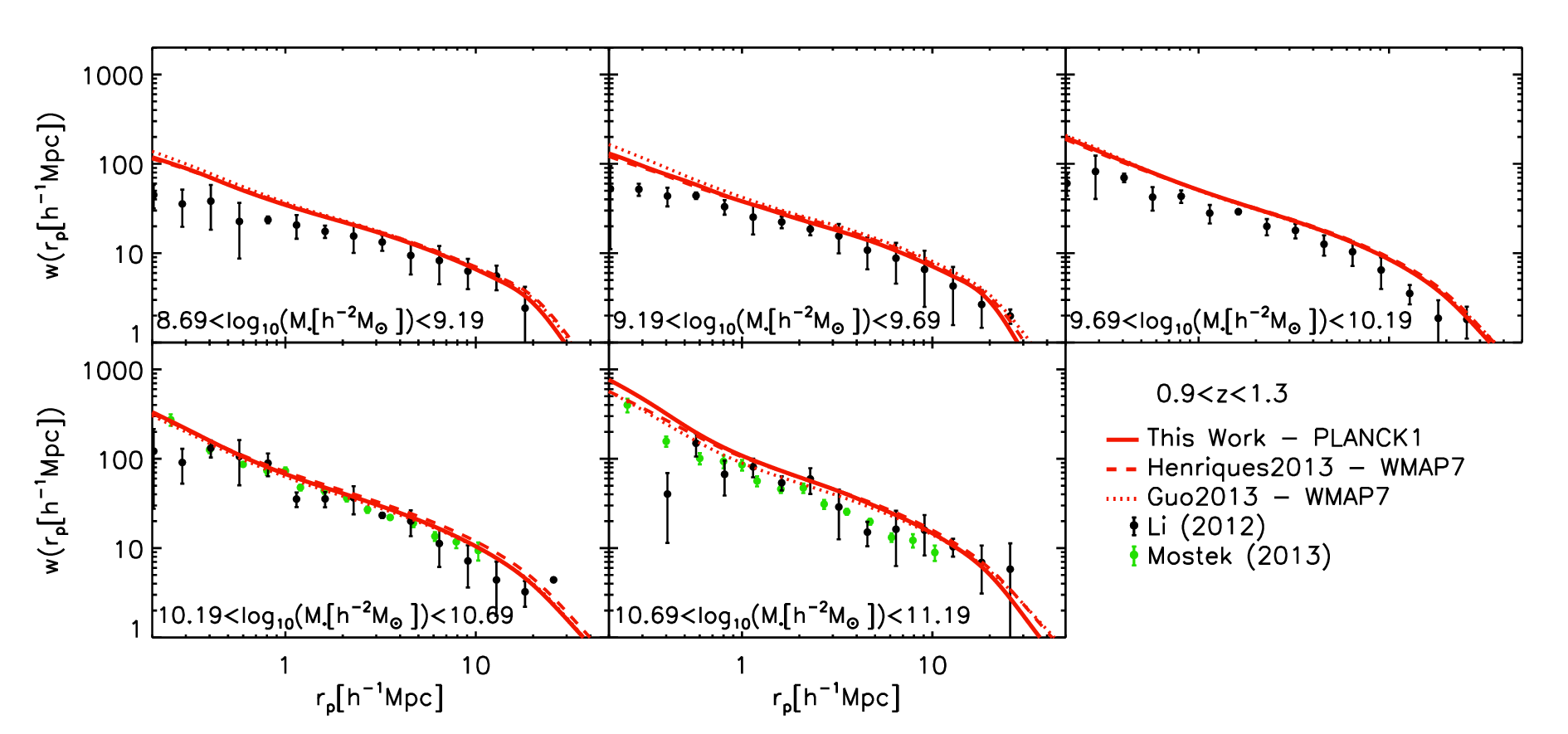}
\caption{Projected 2-point autocorrelations for galaxies in
  five disjoint stellar mass ranges at $z\sim 1$. Results from our new
  model with its best-fit parameter values (solid red lines) are
  compared with model results from \citet{Henriques2013} (dashed red
  lines) and \citet{Guo2011} (dotted red lines) and with observational
  results from DEEP2 (black and green symbols taken from
  \citealt{Li2012} and \citealt{Mostek2013}, respectively). The Millennium-II Simulation is used for
  $\log_{10} (M_*[\Msun]) \leq 9.5$ and the Millennium Simulation for higher
  stellar masses.}
\label{fig:correlation_z1}
\end{figure*}

\section{Galaxy clustering}
\label{sec:clustering}

Galaxy clustering can be measured relatively precisely and in
considerable detail using 2-point auto- and cross-correlations as a
function of scale, redshift and galaxy properties
\citep[e.g.][]{Li2006, Zehavi2011,Mostek2013}. These provide a
stringent test of whether galaxy formation models form the right
galaxies in the right places and at the right times. They are of
particular interest for our current study, since small-scale
correlations are very sensitive to how galaxies populate dark matter
halos, and so to the environmental processes which are our primary
concern.  These produce systematic differences between central and
satellite galaxies, as well as systematic variations of satellite
properties with halo mass and halocentric radius.  

In principle, clustering is also sensitive to the parameters of the
background cosmological model, since these determine the abundance and
clustering of dark matter haloes (e.g. \citealt{Mo2002}).
\citet{Guo2013} showed that the parameter shifts from {\it WMAP1} to
{\it WMAP7} result in rather small shifts in the abundance and
clustering of halos which can easily be compensated by small changes
in galaxy formation parameters. As shown in Paper I, the move to a
{\it Planck} cosmology brings dark halo properties even closer to
those in the original {\it WMAP1} cosmology of the Millennium
Simulations. As a result, any difference in clustering between the model
of this paper and that of \cite{Guo2011} is likely a consequence
of changes in astrophysical modelling, rather than of changes in
cosmology. Indeed, \citet{Henriques2013} found that changes in the
strength of feedback and in the reincorporation time for ejected gas
produce larger changes in how low-mass galaxies populate massive
haloes than any of the cosmological parameter shifts they considered.
Here we will further test the impact of our modified physical modelling 
and of adopting a {\it Planck} cosmology, and we will extend the
analysis to $z=1$.  We will also study clustering differences between
active and passive galaxies. These directly reflect the nature of the
quenching processes which concern us.

\subsection{Projected autocorrelations as a function of stellar mass at low redshift}
\label{subsec:correlation_z0}

In Fig.~\ref{fig:correlation_z0} we plot projected autocorrelation
functions for galaxies in a series of disjoint stellar mass
ranges. Predictions from the models of this paper (solid red lines),
of \citet{Henriques2013} (dashed red lines) and of \citet{Guo2013}
(dotted red lines) are compared with SDSS/DR7 results taken from
\citet{Guo2011}. The first and second models differ primarily in
cosmology, the second and third only in galaxy formation modelling. As
already discussed in \citet{Henriques2013}, the most significant
differences between the \citet{Guo2013} model and the other two are
the enhanced supernova feedback strength and the altered scaling of
the gas reincorporation time. As a result, low-mass galaxies are
significantly less clustered both in \citet{Henriques2013} and in our
new model than in \citet{Guo2013}. This reduction is slightly
compensated by the particular best-fit parameter choice adopted in the
new model, resulting in somewhat better agreement with observation for
$9.0<\log_{10}(M_{\star}[h^{-2}\Msun])<10.0$. Nevertheless,
\citet{Guo2013} (and indeed also \citealt{Guo2011}) still reproduces
observed low-redshift clustering better than our more recent
models in this mass interval. This emphasises the need to use clustering observations
directly as additional constraints in future MCMC analyses if these
are to exploit fully the information content of modern surveys
\citep[see][for a first attempt]{VanDaalen2016}.  The weak impact of
cosmology is evident in the agreement of the present model with
\citet{Henriques2013} on small scales and in the agreement of all
three models on large scales.  This reflects the very small shifts in
cosmological parameters (and hence in structure growth) allowed by
modern CMB data.


\begin{figure*}
\centering
\includegraphics[width=17.9cm]{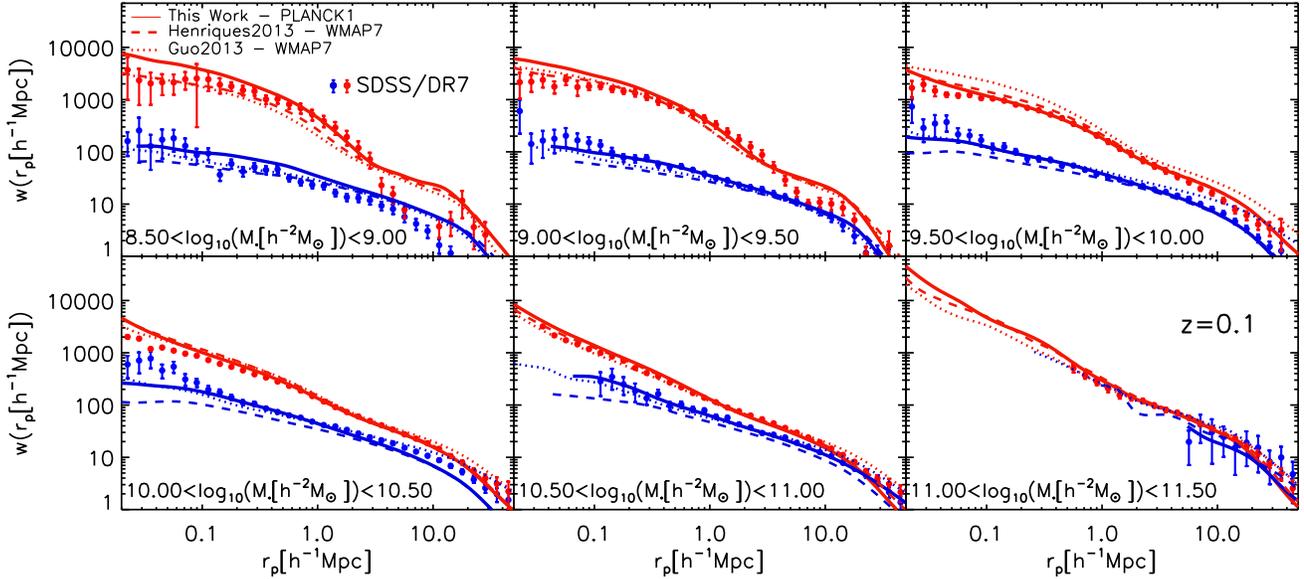}
\caption{Projected 2-point autocorrelations for galaxies in six
  disjoint stellar mass ranges at $z=0.1$ and for blue and red
  galaxies separated according to rest-frame $g-r$ colour. Results
  from our new model with its best-fit parameter values (solid lines)
  are compared with model results from \citet{Henriques2013} (dashed
  lines) and \citet{Guo2011} (dotted lines) and with observational
  results from SDSS/DR7 (symbols, taken from \citealt{Guo2011}). Red
  and blue colours correspond to the passive and active galaxy
  populations, respectively. The Millennium-II Simulation is used for
  $\log_{10} (M_*[\Msun]) \leq 9.5$ and the Millennium Simulation for higher
  stellar masses.}
\label{fig:correlation_bycolor}
\end{figure*}

\subsection{Projected autocorrelations as a function of stellar mass at $z\sim 1$}
\label{subsec:correlation_z1}

Having compared clustering predictions for our new model with
low-redshift observations, we now repeat the exercise at $z\sim 1$,
the earliest time for which there are detailed measurements. This
provides further insight into whether galaxies are correctly
populating the growing dark halo distribution. In
Fig.~\ref{fig:correlation_z1} we compare projected autocorrelation
functions for our current model (solid red lines) and for those of
\citet{Henriques2013} (dashed red lines) and \citet{Guo2013} (dotted
red lines) with two observational analyses of the DEEP2 survey, green
filled circles from \citet{Mostek2013} and black filled circles from
\citet{Li2012}.

The similarity between the three models is striking, holding over
factors of 300 in stellar mass and 1000 in radius.  While the
difference in the treatment of low-mass galaxies between
\citet{Guo2013} and the two more recent models results in different
clustering strengths at $z\sim0$, the three models make almost
identical clustering predictions at $z=1$, despite the fact that the
predicted abundance of low-mass galaxies is significantly reduced at
this redshift in the newer models \citep[see][and Paper
  I]{Henriques2013}.  There is also remarkable agreement both at low
redshift and at $z\sim 1$ between the results of \citet{Henriques2013}
in the {\it WMAP7} cosmology and our new model in its {\it Planck}
cosmology. As noted by \citet{Guo2013}, although there are small
differences in the growth of structure in the two cosmologies, these
are not visible in galaxy clustering once the physical parameters have
been re-adjusted to match the stellar mass function at the relevant
times.  Finally, all models predict clustering at $z\sim 1$ which
appears slightly stronger than observed, particularly for
$\log_{10}(M_{\star}[h^{-2}\Msun])<10.19$ and on small scales. This
may reflect problems with the model's treatment of satellite dynamics
or in our rough approximation to the completeness corrections and 
selection functions used when analysing the observational data.

\subsection{Low-redshift projected autocorrelations as a function of
  stellar mass and colour}
\label{subsec:correlation_bycolor}

In this subsection we compare our new model to observations of the
low-redshift clustering of galaxies split by colour into star-forming
and passive populations (the separation between red and blue galaxies is described in subsection~\ref{subsec:separation}). 
The difference in clustering between red and blue galaxies is very sensitive to the
physics of quenching which is the main concern of
this paper. Quenching by environment will turn satellite galaxies red,
particularly in massive haloes where ram-pressure stripping is effective,
but does not depend strongly on the mass of the satellite. Radio-mode 
AGN quenching, on the other hand, primarily affects the central galaxies 
of relatively massive halos, which themselves have relatively high stellar
mass.  

On large scales galaxy autocorrelations are due to galaxy pairs which
reside in different halos and, except at relatively small stellar
mass, are dominated by pairs in which both galaxies are centrals. As a
result, the colour dependence of autocorrelation strength at given
stellar mass arises partly because red centrals live in more massive
haloes than blue ones \citep[e.g.][and references
  therein]{Mandelbaum2016} and partly because the (subdominant) fraction
of pairs where at least one galaxy is a satellite (and hence typically
lives in a substantially more massive halo) is larger for red galaxies
than for blue ones.

In contrast, small-scale galaxy autocorrelations are dominated by
pairs in which both galaxies reside in the same halo, so that at least
one of them is a satellite. For low-mass galaxies, both galaxies are
satellites in the majority of close pairs. As a result, small-scale
autocorrelations as a function of colour are very sensitive to
environmental quenching.

In Fig.~\ref{fig:correlation_bycolor} the red and blue symbols give
autocorrelation functions estimated separately from the SDSS/DR7 main
galaxy sample for active and passive galaxy samples split according to
the above colour cut. Results are shown for six disjoint ranges in
stellar mass spanning a total range of a factor of 1000. In the two
lowest stellar mass bins the total volume surveyed is relatively small
and the error bars underestimate the uncertainties which are dominated
by cosmic variance. This primarily affects the large-scale
clustering. The curves in each panel give predictions for the same
three models as in Figures~\ref{fig:correlation_z0}
and~\ref{fig:correlation_z1}.  

The variation with colour of these autocorrelation functions is large,
both for the observations and for the models. The difference in
behaviour between the one- and two-halo terms is also very clear,
particularly for the lower stellar masses, where the amplitude of the
one-halo term is well over an order of magnitude higher for red than
for blue galaxies.  On large scale, the red to blue difference is
still substantial (of order a factor of two) except for the highest
stellar masses, where there are very few blue galaxies. 

Given the very large dynamic range in both stellar mass and projected
radius which is plotted here, the models agree remarkably well
(although not perfectly) with the SDSS data, suggesting that the
interplay of internal and environmental effects is being relatively
well captured as a function of stellar mass.  Further improvement will
require a more rigorous characterisation of the uncertainties in the
observed autocorrelations, and the use of the clustering measurements
as additional constraints in the MCMC exploration of model parameter
space (e.g. by extending the methods in \citealt{VanDaalen2016}).

\begin{figure}
\centering
\includegraphics[width=8.7cm]{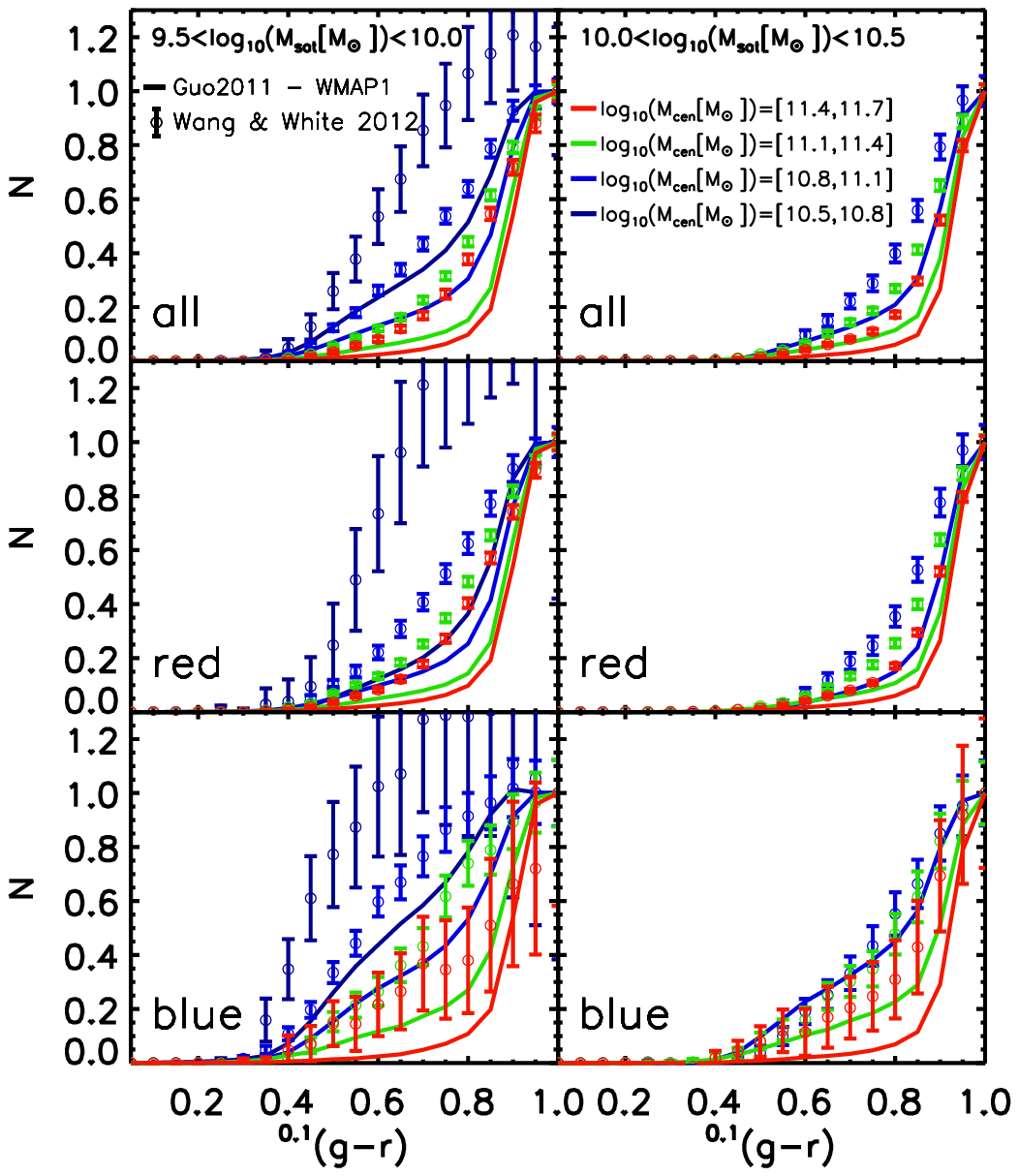}
\caption{Cumulative $^{0.1}(g-r)$ colour distributions for satellites
  projected within 300\,kpc of their primaries. The \citet{Guo2011}
  model (solid lines) is compared with SDSS observations (symbols with
  error bars) for two different satellite stellar mass ranges (the
  left and right columns) around primaries in four different stellar
  mass ranges (indicated by the colours). In each case, results are
  given for the primary sample as a whole (in the top row) and for the
  sample split into red (middle row) and blue (lower row) primaries.
  The observational data in this plot are taken from Fig.12 of
  \citet{WangWhite2012} while the model results are taken from their
  Fig.13.}
\label{fig:confor_G11}
\end{figure}

\section{Conformity}
\label{sec:conformity}

In previous sections we explored how well the observed trends of
quenching with stellar mass and environment are reproduced by our new
model. Although treated as separate processes, AGN quenching and
environmental quenching are clearly connected in the model. Hot gas 
masses are larger in more massive halos, leading to enhanced black hole 
accretion and AGN feedback in denser environments. In addition, such 
environments promote black hole growth by enhancing both mergers with 
smaller black holes and merger-stimulated accretion of cold gas. At the 
same time, larger halo and hot gas masses boost
environmental quenching through tidal and ram-pressure stripping. As a
result, one may expect that quenched central galaxies will typically
have more massive halos and more massive hot gas atmospheres than blue
centrals of similar stellar mass, and that this will cause a larger
fraction of their satellites to be quenched. \citet{WangWhite2012}
showed that this effect is indeed present in the \citet{Guo2011} model
and compared its strength to the effects they found in SDSS data,
demonstrating that all the observed trends with satellite and primary
stellar mass are reproduced qualitatively by the model but that its
satellites are systematically too red.

\begin{figure}
\centering
\includegraphics[width=8.7cm]{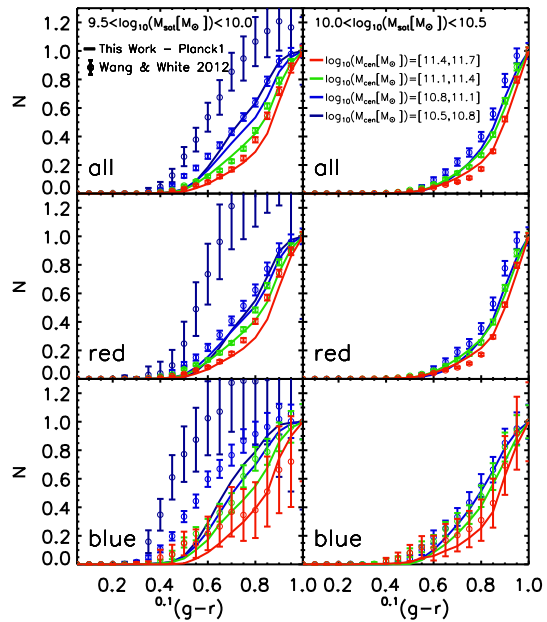}
\caption{As Fig.~\ref{fig:confor_G11} but now for our new model rather
  than that of \citet{Guo2011}.  Note that all the trends seen in the
  data and in the earlier model are also present in the new model, but
  that the satellite colour distributions are systematically shifted
  bluewards, bringing them into reasonable agreement with the SDSS
  data for all except low-mass satellites of the low-mass 
  primaries.}
\label{fig:confor_H14}
\end{figure}

The effect that passive centrals tend to have passive satellites and
vice-versa was first noticed in SDSS groups by
\citet{Weinmann2006b} who designated it ``galactic conformity''.  A
number of studies have confirmed the strength of this tendency and
investigated its variation with primary and satellite properties, with
halo mass, with halocentric distance and with redshift
\citep{Ann2008,Kauffmann2010,Prescott2011,WangWhite2012,Kauffmann2013,
  Phillips2014,Hartley2015,Knobel2015}. By counting satellite
candidates to $r=21$ in the SDSS/DR8 photometric catalogues around
isolated central galaxies brighter than $r=16.7$,
\citet{WangWhite2012} were able to measure conformity over a wide
range of primary and satellite stellar masses and we shall compare our
current model to their data in this section. We therefore define model
primaries in a manner which closely reflects their observational
criteria.

Specifically, we select isolated galaxies as those for which every
companion projected at $r_p < 0.5$\,Mpc and with redshift difference
$|\Delta z| < 1000\; \rm{km}\,\rm{s}^{-1}$ is at least a magnitude
fainter in the SDSS $r-$band, and for which no companion within $r_p <
1.0$\,Mpc and $|\Delta z| < 1000\; \rm{km}\,\rm{s}^{-1}$ is
brighter. The simulation boxes are projected in three orthogonal
directions, parallel to their x, y and z axes, and redshifts are
assigned based on the ``line-of-sight distance'' and peculiar velocity
of a galaxy. For a set of primaries, defined as isolated galaxies with
stellar mass and colour lying in predetermined ranges, we then
identify the set of potential satellites as all objects at least a
magnitude fainter than, and projected within 300~kpc of, some
primary. This set will include foreground and background objects
unassociated with the primaries. We correct for these using the mean
surface density of galaxies as a function of absolute magnitude and
colour for the simulation as a whole (see \cite{WangWhite2012} for
details).

Fig.~\ref{fig:confor_G11} shows the comparison of SDSS observational
results to the model of \cite{Guo2011} that was already presented in
\citet{WangWhite2012} (their Figs. 12 and 13), while
Fig.~\ref{fig:confor_H14} is identical except that the data are now
compared with our new model. In each panel, coloured symbols show the
cumulative distribution of $^{0.1}(g-r)$ colour for SDSS satellites,
with error bars derived by boot-strap resampling of the primary
sample, while the coloured curves refer to the models and have
negligible statistical uncertainty. In order to have a fair comparison 
with observations, model colours at $z=0$ are 
converted to $^{0.1}(g-r)$ using the relations from \citet{Blanton2007}:
$^{0.1}g=g+0.3113+0.4620\times((g-r)-0.6102)$ and 
$^{0.1}r=g-0.4075-0.8577\times((g-r)-0.6102)$.
Each colour corresponds to a
disjoint primary stellar mass range, as indicated in the legend. The
two columns of panels refer to the two disjoint ranges of satellite
stellar mass indicated by labels above each column. The three rows
refer respectively to all primaries regardless of colour (top row) to
red primaries (middle row) and to blue primaries (bottom row). 
The details of the separation between red and blue galaxies 
are given in subsection~\ref{subsec:separation}.

Several strong trends are visible in these plots, both in the SDSS
data and in the two models. More massive primary galaxies have
systematically redder satellites (from left to right the curves and
symbols go from dark blue to red in every panel of
Figs.~\ref{fig:confor_G11} and~\ref{fig:confor_H14}). More massive
satellites tend to be redder, at least for the two mass ranges
considered here (curves and symbols shift rightward moving from the
left panel to the right panel in each row). For all four primary mass
ranges and for both satellite mass ranges, red primaries have
systematically redder satellites than blue primaries of the same
stellar mass (the curves and symbols in every middle row panel lie to
the right of the corresponding curves and symbols in the panel
immediately below it). This last trend is the galactic conformity
highlighted by \citet{Weinmann2006b}. For satellites projected within
300~kpc, it is clearly present both in the observations and in the
models over a range of a factor of 20 in primary stellar mass and a
factor of 10 in satellite stellar mass. 

As pointed out by \cite{WangWhite2012}, in every panel of
Fig.~\ref{fig:confor_G11} the model curves lie well to the right of
the corresponding symbols, showing that satellites are systematically
too red in the \cite{Guo2011} model. As can be seen in
Fig.~\ref{fig:confor_H14}, the situation is substantially improved in
our new model where satellites are bluer in all cases and large
discrepancies remain only for low-mass satellites of low-mass
primaries.  This is a result of the weaker impact of ram-pressure
stripping, combined with the lowered threshold for star formation
which allows satellites to keep forming stars for longer. It seems,
nevertheless, that environmental effects are still too strong for
low-mass satellites of low-mass primaries. As discussed in
\citet{Wang2014}, where an early version of the current model was
found to produce radial distributions of blue satellites which are
significantly less centrally concentrated than observed in the SDSS
data, a possible solution may be to balance environmental effects with
interaction-induced starbursts. These would enhance star formation in
satellites close to orbital pericentre.

\citet{WangWhite2012} argued that, in the model of \cite{Guo2011},
conformity arises because red primaries have more massive haloes and
more massive hot gas atmospheres than blue primaries of the same
stellar mass, resulting in stronger environmental effects on their
satellites. These effects are also present in our new model and may
explain the strong quenching conformity between centrals and
satellites found by \citet{Knobel2015} in their own analysis of SDSS
data. On the other hand, \citet{Kauffmann2013} found conformity
effects to persist out to projected distances of order 4~Mpc,
considerably larger than the expected halo size of the central
galaxies. They showed that although an effect of the same sign is
present in the \cite{Guo2011} model, it is much weaker than
observed. \cite{Hearin2015,Hearin2016} show that small effects of the
size seen in the model can be explained by the tendency of pairs of
dark halos at these separations to have correlated assembly histories.
The origin of the strong effect seen by \citet{Kauffmann2013} remains
uncertain, however.  At these separations, conformity is actually
somewhat weaker in our new model than in that of
\cite{Guo2011}. \citet{Kauffmann2015} presents some evidence suggesting
that a nongravitational process related to early AGN activity may be
responsible for the observed effect.


\section{Summary}
\label{sec:conclusions}

The current version of the Munich galaxy formation model, introduced
in \cite{Henriques2015}, applies the publicly
available {\small L-GALAXIES} semi-analytic modelling code\footnote{http://galformod.mpa-garching.mpg.de/public/LGalaxies/} to the Millennium and
Millennium-II Simulations after scaling them to the Planck cosmology
following \cite{Angulo2015}. The parameters of the galaxy formation
model were adjusted to fit the observed variation of galaxy abundance
and passive fraction as functions of stellar mass over the redshift
range $0\leq z\leq 3$. A good fit required significant modification of
the physical assumptions underlying previous public versions of the
Munich models \citep{Guo2011,Guo2013}. In particular, to ensure that
low-mass galaxies form most of their stars at $z< 1.5$ and are
predominantly star-forming at $z=0$, the ejection of material by SN
feedback had to be increased and the reincorporation of ejected
material had to be enhanced at late times
\citep[see][]{Henriques2013}.  These changes increased star-formation
rates at low redshift in low-mass central galaxies, but did not
prevent such galaxies from turning red too quickly when they became
satellites. The current model, in addition, reduces the cold gas
density threshold for star formation and assumes ram-pressure
stripping to be negligible in low-mass groups, thereby significantly
increasing the fraction of low-redshift satellites with ongoing star
formation.  With these assumptions, intermediate mass galaxies
continue to grow in mass at least down to $z\sim 1$. To ensure that
they are nevertheless predominantly quenched by $z=0$, it was
necessary to change the scaling of AGN feedback to make it more
efficient at late times.

In Paper I we showed that this new model agrees well at all redshifts
with the observed stellar mass functions and passive fractions used to
calibrate it. In \cite{Shamshiri2015} we developed new methods which
retain detailed star formation histories for all galaxies at all
redshifts. In combination with the population synthesis codes, these
can be used (in post-processing) to obtain integrated spectra and
magnitudes in arbitrary optical/near-infrared passbands for all
objects. In \cite{Clay2015} we compared observed and modelled
abundances and photometric properties at $4\leq z \leq 7$, well beyond
the redshift range used for calibration.  In the current paper, we
concentrate on the spatial distribution of galaxies in order to gauge how
well our model represents the observed quenching of galaxies as a
function of mass, environment and epoch. Quenching in the model is
driven by AGN feedback, which suppresses gas accretion onto massive
galaxies at the centres of hot gas halos, and by environmental effects
on satellite galaxies, namely the suppression of cosmological infall,
the tidal stripping of haloes, and ram-pressure stripping in galaxy
clusters. These processes interact in a complex way and it is
important to replicate observational procedures closely in order to
understand whether the observed trends are reproduced.

We began this paper by showing that when split by colour into
star-forming and passive subsets the stellar mass functions of
galaxies are a good fit to available observations over the full range
$0\leq z \leq 3$, and as a result they reproduce the phenomenology
noted by \cite{Peng2010} over the more restricted range $0\leq z \leq
1$ from their analysis of SDSS and COSMOS data. Namely that the SMF of
star-forming galaxies evolves little with redshift and has a steep
faint-end slope, while the SMF of passive galaxies has two components,
one (corresponding to satellites) which has the same shape as that of
star-forming galaxies and evolves equally slowly but at lower
amplitude, and a second (corresponding to centrals) which has a much
shallower low-mass slope and grows substantially in amplitude at low
redshift. Even at fixed environment density (as estimated from the
distance to the fifth nearest neighbour) passive fractions in the
model vary with stellar mass at given redshift in a way which
resembles that found by \cite{Peng2010}. As a result, if we infer
``mass quenching efficiencies'' from our model in the same way as they
do from the data, we find results which are in qualitative agreement
with theirs over the full range $0<z<1$, although the observationally
inferred dependence on stellar mass is steeper than in the model. This
difference is also seen in a comparison with SDSS data as compiled by
\cite{Wetzel2012} who split observed galaxies into ``centrals'' and
``satellites'' and used inferred halo mass as a measure of satellite
environment. Another important outcome of our AGN-feedback-dependent 
quenching of massive galaxies is that it naturally explains why these are observed to 
host more massive black holes when they are quenched than when they are 
still star-forming \citep{Terrazas2016b}.

\cite{Wetzel2012} explored the environment dependence of quenching in
more detail by compiling SSFR
histograms and radial profiles of passive fraction for satellite
galaxies split according to stellar mass and halo mass. Our current
model matches the SSFR histograms quite well for all stellar and halo
masses, but is less successful in matching the radial profiles, which
are significantly steeper than in the SDSS data. It appears that the
radial dependence of quenching is overly strong in the model. On the
other hand, the dependence of quenching on halo mass appears to be
somewhat weaker in the model than in the \cite{Wetzel2012} compilation
of SDSS data, although the qualitative agreement is quite good. For
the \cite{Peng2010} definition of environment density, this problem
shows up as a stronger dependence of quenched fraction on environment
than is observed. The evolution of ``environmental quenching
efficiency'' in our model as defined by them is qualitatively similar to what they
found, albeit with a stronger dependence on redshift.

The two-point statistics of galaxy clustering as a function of stellar
mass/luminosity and colour/SSFR provide a well established route to
studying the environment dependence of galaxy formation and evolution.
Observational determinations were already compared to predecessors of
our current model in \cite{Springel2005} and \cite{Guo2011}. The
latter paper found agreement with low-redshift SDSS data to be good
for massive galaxies and on large scales, but uncovered significant
discrepancies at small scale for low-mass galaxies, particularly for
passive systems. This problem reflected an overly large
passive fraction among satellite galaxies. The current model has
decreased these discrepancies significantly, but to fully explore how
clustering data depend on galaxy formation physics, it will be
necessary to characterise better the observational uncertainties of
the clustering measurements and to use them explicitly as constraints
in the MCMC sampling of the galaxy formation parameter space
\citep[see][for a first attempt in this direction]{VanDaalen2016}.  Two
point clustering data have recently become available at $z\sim 1$ and
our current model agrees quite well with the autocorrelations as a
function of stellar mass from DEEP2.
 
A particular aspect of galaxy two-point clustering statistics which
has received considerable recent attention is the ``conformity''
signal first pointed out by \cite{Weinmann2006b}. Satellites of
passive central galaxies are more likely themselves to be passive than
satellites of star-forming central galaxies of the same stellar
mass. \cite{Kauffmann2013} pointed out that this signal extends out to
companions at distances of several Mpc. These are too distant to be
satellites as normally defined. We compared our model to the detailed
information on satellite conformity compiled by \cite{WangWhite2012}
from SDSS data. As was the case for the \cite{Guo2011} model to which
\cite{WangWhite2012} compared, all the trends seen in the observations
are qualitatively reproduced by the model.  However, whereas the
earlier model produced satellites that were systematically too red at
all primary and satellite stellar masses, the distributions have
shifted blue-ward in the new model and are in approximate agreement
with the data for all but low-mass satellites of low-mass primaries.

As discussed by \cite{WangWhite2012}, strong satellite-central
conformity arises in the model because passive centrals tend to have
more massive halos and more massive hot gas atmospheres than
star-forming centrals of the same mass. The first effect is due to
fact that the haloes of star-forming centrals grow in parallel with
the galaxy up to the present day, whereas the haloes of passive
centrals continue to grow in mass after star formation has stopped.
The second effect arises because the strength of AGN feedback
increases with the mass of hot gas. These effects result in conformity
because satellites are more likely to be quenched in more massive
haloes with a denser gaseous atmosphere. Thus, halo mass and hot gas mass are acting as
halo-wide ``hidden variables'' of the kind which \cite{Knobel2015}
inferred to be required in their own study of conformity. On the other
hand, our new model is no better at reproducing the large-scale and
strong apparent ``two-halo conformity'' seen in the SDSS data than was
the \cite{Guo2011} model used as comparison in the original
\cite{Kauffmann2013} analysis.

Overall, the publicly available galaxy
formation simulation presented in Paper I and further analysed here
gives a good representation of the abundance and star formation
distributions of galaxies over the full observed stellar mass range
out to $z\sim 7$, as well as of galaxy clustering and the correlation
of galaxy properties with environment out to $z\sim 1$. The clearest remaining
discrepancies are the overly strong dependence of satellite quenching
on halocentric radius, the lack of a strong two-halo conformity
signal and the inability of our AGN feedback model to quench
star formation in all galaxies with $\log_{10} (M_*[\Msun]) \ge 11.0$ 
at low redshift ($20\%$ of these have ongoing star formation).
This last problem seems to be alleviated if we allow bar-driven accretion 
onto black holes. The advantages of our semi-analytic simulation and others 
like it over purely phenomenological models for the evolution of the galaxy
population are that it automatically ensures
consistency between the populations observed at different redshifts,
and it allows observed trends to be interpreted directly in terms of
physical processes. Its advantages over hydrodynamical simulations are 
that much larger volumes can be simulated, and that
systematic exploration of the high-dimensional parameter space of
galaxy formation is nevertheless feasible. Future work will
undoubtedly bring further discrepancies to light and confirm the
significance of those already identified. Exploring their origin
should give additional insight into the astrophysics of galaxy
formation and evolution.

\section*{Acknowledgements}
This work used the DiRAC Data Centric system at Durham University,
operated by the Institute for Computational Cosmology on behalf of the
STFC DiRAC HPC Facility (www.dirac.ac.uk). This equipment was funded
by BIS National E-infrastructure capital grant ST/K00042X/1, STFC
capital grant ST/H008519/1, and STFC DiRAC Operations grant
ST/K003267/1 and Durham University. DiRAC is part of the National
E-Infrastructure. BMBH (ORCID 0000-0002-1392-489X) acknowledges support from a Zwicky Prize fellowship. 
The work of BMBH, SDMW and GL was supported by Advanced
Grant 246797 ``GALFORMOD'' from the European Research Council. PAT (ORCID 0000-0001-6888-6483) 
acknowledges support from the Science and Technology Facilities Council (grant number ST/L000652/1). 
REA acknowledges support from AYA2015-66211-C2-2. 
GQ acknowledges support from the NSFC grant (Nos. 11573033, 11622325), 
the Recruitment Program of Global Youth Experts of China, the NAOC grant (Y434011V01), 
MPG partner Group family, and a Royal Society Newton Advanced Fellowship.  
The authors thank Ivan Baldry,
Olivier Ilbert, Alexander Karim, Cheng Li, Danilo Marchesini and Adam
Muzzin for providing their observational data, Eric Bell, Jarle
Brinchmann, Guinevere Kauffmann, Bryan Terrazas, Stephen Wilkins and
Rob Yates for useful discussions and Claudia Maraston, Gustavo Bruzual
and Stephan Charlot for providing their stellar populations synthesis
models.

\bibliographystyle{mn2e} \bibliography{paper_HWT14b}

\appendix


\label{lastpage}

\end{document}